# Numerical simulation of time delay interferometry for new LISA, TAIJI and other LISA-like missions

Gang Wang[1,2] and Wei-Tou Ni[3,4]

[1]INFN, Gran Sasso Science Institute, I-67100 L'Aquila, Italy
[2]INFN, Sezione di Pisa, Edificio C, Largo Bruno Pontecorvo, 3, I-56127, Pisa, Italy
[3]Center for Gravitation and Cosmology (CGC), Department of Physics, National Tsing Hua University, No. 101, Kuang Fu II Rd., Hsinchu, Taiwan, 300 ROC
[4]Kavli Institute for Theoretical Physics China, CAS, Beijing 100190, P. R. China

E-mail: gwanggw@gmail.com, weitou@gmail.com

**Abstract**. With aLIGO's discovery of the gravitational wave (GW) sources, we were ushered into the age of observational GW astronomy. The success of LISA Pathfinder in demonstrating the LISA drag-free requirement paved the road of using space missions for detecting low-frequency and middle-frequency GWs. The new LISA GW mission proposes to use arm length of 2.5 Gm (1 Gm = $10^6$ km). The TAIJI GW mission proposes to use arm length of 3 Gm. In order to attain the requisite sensitivity, laser frequency noise must be suppressed to below the secondary noises such as the optical path noise, acceleration noise etc. In previous papers, we have performed the numerical simulation of the time delay interferometry (TDI) for original LISA, ASTROD-GW and eLISA together with a LISA-type mission with a nominal arm length of 2 Gm using the CGC 2.7/CGC2.7.1 ephemeris framework. In this paper, we follow the same procedure to simulate the time delay interferometry numerically for the new LISA mission and the TAIJI mission together with LISA-like missions of arm length 1 Gm, 2 Gm, 4 Gm, 5 Gm and 6 Gm. To do this, we work out a set of 2200-day (6-year) optimized mission orbits of each mission starting at March 22, 2028 using the CGC 2.7.1 ephemeris framework. We then use numerical method to calculate the residual optical path differences of the first-generation TDI configurations --- Michelson X, Y & Z; Sagnac $\alpha$, $\beta$ & $\gamma$; Relay U, V & W; Beacon P, Q & R; Monitor E, F & G --- and the second generation TDI configurations --- 2-arm type, [*ab*, *ba*]'s (*a*, optical path from S/C 1 to S/C 2 and back to S/C 1; *b*, optical path from S/C 1 to S/C 3 and back to S/C 1); [*aabb*, *bbaa*]'s; [*abab*, *baba*]'s; [*abba*, *baab*]'s; Sagnac-type $\alpha2$, $\beta2$ & $\gamma2$. The resulting optical path differences of the second-generation TDI calculated for new LISA, TAIJI, and LISA-like missions or arm length 1, 2, 4, 5 & 6 Gm are well below their respective limits which the laser frequency noise is required to be suppressed. However, for of the first generation X, Y, and Z TDI configurations, the original requirements need to be relaxed by 3 to 30 fold to be satisfied. For the new LISA and TAIJI, about one order of magnitude relaxation would be good and recommended; this could be borne on the laser stability requirement in view of recent progress in laser stability. Compared with X, Y and Z, the X+Y+Z configuration does have a good cancellation of path length differences and could serve as a null string detection check. We compile and compare the resulting differences of various TDI configurations due to the different arm lengths for various LISA-like mission proposals and for the ASTROD-GW mission proposal.

## 1. Introduction and summary

GW detection has been a focused subject of research for some time. Detection efforts in all GW frequency bands from Hubble frequency band (1 aHz-10 fHz) to ultra-high frequency band (over 1 THz) are pursued vigorously (See, e.g. Kuroda *et al* 2015). With the announcement of LIGO direct Gravitational Wave (GW) detection (Abbott *et al* 2016a, 2016b, 2017), we are fully ushered into the age of GW astronomy.

*1.1. Space laser-interferometric GW detectors*

Space laser-interferometric GW detectors operate in the low frequency band (100 nHz–0.1 Hz) and the middle frequency band (0.1 Hz–10 Hz). The scientific goals are to detect following GW sources in these bands: (i) Massive black holes (BHs); (ii) Extreme mass ratio inspirals; (iii) Intermediate mass black holes; (iv) Compact binaries; (v) Relic GWs, and to use observation and measurement of these sources to study the co-evolution of massive BHs with galaxies, to anticipate binary merging GW events in the high frequency band (10 Hz–100 kHz) for Earth-based GW detection, to test relativistic gravity, to determine cosmological parameters, and to study dark energy equation of state. Space GW detectors may provide much higher S/N ratio for GW detection than Earth-based detectors.

Interferometric GW detection in space basically measures the difference in the distances traveled through two routes of laser links among S/C (or celestial bodies) as GWs come by. The S/C (or celestial bodies) must be in geodesic motion (or such motion can be deduced). The distance measurement must be ultra-sensitive as the GWs are weak. Therefore, we must use drag-free technology to guarantee the geodesic motion and laser stabilization to the required level of measurement for achieving the scientific goals. Due to long distance between spacecraft required for low frequency (100 nHz–0.1 Hz) and middle frequency (0.1 Hz–10 Hz) GW measurement, we must use appropriate amplification method between laser links. This is achieved by either homodyne or heterodyne optical phase locking the local oscillator to the incoming weak light at the received link.

LISA (Laser Interferometer Space Antenna) Pathfinder (Armano *et al* 2016), launched on 3 December 2015, has achieved not only the drag-free requirement of this technology demonstration mission, but also completely met the more stringent LISA drag-free demand (Amaro-Seone *et al* 2017). In short, LISA Pathfinder has successfully demonstrated the drag-free technology for space detection of GWs.

At National Tsing Hua University, 2 pW weak-light homodyne phase-locking with 0.2 mW local oscillator has been demonstrated (Liao *et al* 2002a, 2002b). In JPL (Jet Propulsion Laboratory), Dick *et al* (2008) have achieved offset phase locking local oscillator to 40 fW incoming laser light. Recently, Gerberding *et al*. (2013) and Francis *et al*. (2014) have phase-locked and tracked a 3.5 pW weak light signal and a 30 fW weak light signal respectively at reduced cycle slipping rate. For LISA, 85 pW weak-light phase locking is required. For ASTROD-GW, 100 fW weak-light phase locking is required. Hence, the weak-light power requirement is demonstrated. In the future, frequency-tracking, modulation-demodulation and coding-decoding needs to be well-developed to make it a mature technology. This is also important for CW (Continuous Wave) deep space optical communication (e.g. Dick *et al* 2008).

To reach the measurement sensitivity goal for space detection, we need to suppress spurious noise below the aimed sensitivity level. This requires us to reduce the laser noise as much as possible. The drag-free technology is now demonstrated by LISA Pathfinder. Reducing laser noise requires laser stabilization. However, the best laser stabilization alone at present is not enough for the required strain sensitivity of the order of $10^{-21}$. To lessen the laser noise requirement, TDI came to rescue.

*1.2. Time delay interferometry (TDI)*

For space laser-interferometric GW antenna, the arm lengths vary according to solar system orbit dynamics. In order to attain the requisite sensitivity, laser frequency noise must be suppressed below the secondary noises such as the optical path noise, acceleration noise etc. For suppressing laser frequency noise, it is necessary to use TDI in the analysis to match the optical path length of different beam paths closely. The better match of the optical path lengths are, the better cancellation of the laser frequency noise and the easier to achieve the requisite sensitivity. In case of exact match, the laser frequency noise is fully canceled, as in the original Michelson interferometer.

Except DECIGO (Kawamura *et al*. 2006, 2011) whose configuration with arm length feedback control is basically like the ground GW interferometric detectors, all other laser-interferometric antennas for space detection of GWs have their arm lengths vary with time geodetically (in free fall) according to orbit dynamics. The TDI technique can be used to suppress the laser frequency noise. The basic principle of TDI is to use two different optical paths but whose optical path lengths are nearly equal, and follow them in different/opposite order. This operation suppresses the laser frequency noise if the two paths compared are close enough in optical path length (time travelled).

The TDI was first used in the study of ASTROD mission concept in the 1990s (Ni *et al* 1997, Ni

1997). The following two TDI configurations were used during the study of ASTROD interferometry and the path length differences were numerically obtained using Newtonian dynamics.

These two TDI configurations are the unequal arm Michelson TDI configuration and the Sagnac TDI configuration for three spacecraft formation flight. The principle is to have two split laser beams to go to Paths 1 and 2 and interfere at their end path. For unequal arm Michelson TDI configuration, one laser beam starts from spacecraft 1 (S/C1) directed to and received by spacecraft 2 (S/C2), and optical phase locking the local laser in S/C2; the phase locked laser beam is then directed to and received by S/C1, and optical phase locking another local laser in S/C1; and so on following Path 1 to return to S/C1:

$$\text{Path 1: S/C1} \rightarrow \text{S/C2} \rightarrow \text{S/C1} \rightarrow \text{S/C3} \rightarrow \text{S/C1}. \qquad (1.1)$$

The second laser beam starts from S/C1 also, but follows Path 2 route:

$$\text{Path 2: S/C1} \rightarrow \text{S/C3} \rightarrow \text{S/C1} \rightarrow \text{S/C2} \rightarrow \text{S/C1}, \qquad (1.2)$$

to return to S/C1 and to interfere coherently with the first beam. If the two paths has exactly the same optical path length, the laser frequency noises cancel out; if the optical path length difference of the two paths are small, the laser frequency noises cancel to a large extent. In the Sagnac TDI configuration, the two paths are:

$$\begin{aligned}\text{Path 1: S/C1} \rightarrow \text{S/C2} \rightarrow \text{S/C3} \rightarrow \text{S/C1},\\ \text{Path 2: S/C1} \rightarrow \text{S/C3} \rightarrow \text{S/C2} \rightarrow \text{S/C1}.\end{aligned} \qquad (1.3)$$

Since then we have performed the numerical simulation of the time delay interferometry for ASTROD-GW with no inclination (Wang and Ni, 2012, 2013b), LISA (Dhurandhar, Ni and Wang, 2013), eLISA/NGO (Wang and Ni, 2013a), LISA-type with 2 Gm (1 Gm = $10^9$ m = $10^6$ km) arm length (Wang and Ni, 2013a), and ASTROD-GW with inclination (Wang and Ni, 2015).

TDI has been worked out for LISA much more thoroughly on various aspects since 1999 (Armstrong, Estabrook and Tinto, 1999; Tinto and Dhurandhar, 2014). First-generation and second-generation TDIs are proposed. In the first-generation TDIs, static situations are considered, while in the second generation TDIs, motions are compensated to certain degrees. The two configurations considered above are first-generation TDI configurations in the sense of Armstrong *et al* (1999). For a thorough discussion on the generations and other aspects of TDI, we refer the readers to the excellent review of Tinto and Dhurandhar (2014).

*1.2.1.* X, Y, Z, X+Y+Z, *Sagnac and other first-generation TDI*s

The unequal arm Michelson TDI starting from S/C1 is frequently denoted by the symbol *X*. That starting from S/C2 with (123) permutation in spacecraft is frequently denoted by the symbol *Y* and the one starting from spacecraft 3 by the symbol *Z*. The Sagnac configuration is frequently denoted by the symbol *α*, those with successive permutation(s) by the symbols *β* and *γ*. We shall adopt this notation first used in Armstrong *et al*. (1999).

For the numerical evaluation, we take a common receiving time epoch for both beams; the results would be very close to each other numerically if we take the same start time epoch and calculate the path differences. This way, we can start from Path 1 and propagate the laser light to the end of Path 1 using ephemeris framework, and at the end point of Path 1, evolve laser light back in time along the reversed Path 2 to find the difference in the optical path length. In the TDI, we actually compare the laser signal this way. The results of this calculation of various first-generation TDI orbit configurations are shown in figures 7-13 in section 3 and figures 6-8 in the supplement, and their min, max and rms path length differences for a period of 2200 days are compiled in table 1 of this section; those for second-generation are shown in Fig.'s 14-19 in section 4 and the min, max and rms path length differences for a period of 2200 days are compiled in table 2. For ease to denote the second generation TDIs, we refer to the path S/C1 → S/C2 → S/C1 as *a* and the path S/C1 → S/C3 → S/C1 as *b* as done in Dhurandhar *et al*. (2010, 2013), and Wang and Ni (2013a,b; 2015). Hence the difference Δ*L* between Path 1 and Path 2 for the unequal-arm Michelson X can be denoted as $ab - ba \equiv [a, b]$. To extend this notation to the cyclic permuted paths, we refer to the path S/C2 → S/C3 → S/C2 as *c*, the path S/C2 → S/C1 → S/C2 as *d*, the path S/C3 → S/C1 → S/C3 as *e*, and the path S/C3 → S/C2 → S/C3 as *f*.

The 1st-generation TDIs include Sagnac ($\alpha$, $\beta$, $\gamma$), Unequal-arm Michelson (X, Y, Z), Relay (U, V, W), Beacon (P, Q, R), and Monitor (E, F, G) configurations. The geometric representation of U, P, and E, TDI configurations according to Vallisneri (2005) is shown in Fig. 1. Each type has two other configurations based on different initial points of spacecraft; they can be readily figured out by permutation.

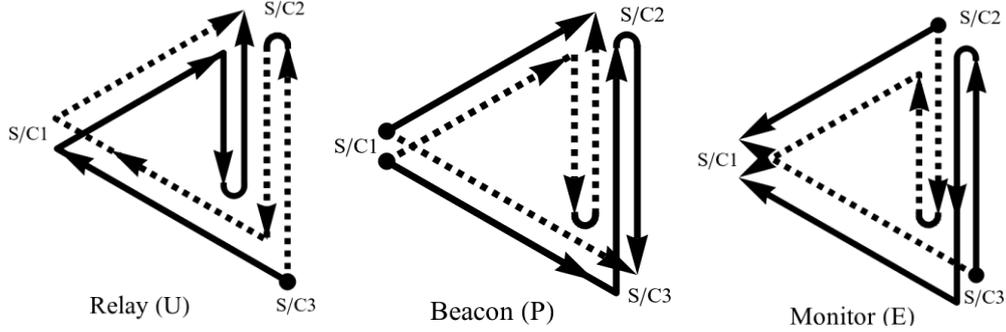

**Fig. 1.** Interference paths of the 1st-generation time-delay interferometry for Relay (U), Beacon (P) and Monitor E.

*1.2.2. Second generation TDIs*

There are many second generation TDI configurations. In this paper we calculated the TDI path length differences for the second-generation TDIs with $n = 1$ and $n = 2$ in the case of one detector with two arms obtained by Dhurandhar *et al.* (2010). These configurations for S/C1 as the start are listed as follows:

(I) $n = 1$, [*ab, ba*] (= *abba* – *baab*),
(II) $n = 2$, [$a^2b^2$, $b^2a^2$], [*abab, baba*]; [$ab^2a$, $ba^2b$].

For S/C 2 and S/C 3 as the starts, the TDI configurations are respectively:

(III) $n = 1$, [*cd, dc*],
(IV) $n = 2$, [$c^2d^2$, $d^2c^2$], [*cdcd, dcdc*]; [$cd^2c$, $dc^2d$],
(V) $n = 1$, [*ef, fe*],
(VI) $n = 2$, [$e^2f^2$, $f^2e^2$], [*efef, fefe*]; [$ef^2e$, $fe^2f$].

A second generation TDI involves 3 arms is the following Sagnac-type TDI. From first-generation Sagnac-$\alpha$ configuration, we add another Sagnac-$\alpha$ in reverse order to get a new interferometry path as follows:

$$\text{Path 1: S/C1} \to \text{S/C2} \to \text{S/C3} \to \text{S/C1} \to \text{S/C3} \to \text{S/C2} \to \text{S/C1},$$
$$\text{Path 2: S/C1} \to \text{S/C3} \to \text{S/C2} \to \text{S/C1} \to \text{S/C2} \to \text{S/C3} \to \text{S/C1}. \quad (1.4)$$

We tag this interferometry configuration as $\alpha 2$ or Sagnac-$\alpha 2$. With cyclic permutations, we obtain other 2 interferometry configurations $\beta 2$ and $\gamma 2$.

*1.3. Orbit configuration for various mission proposals*

In a recent review on the GW detection in space, we have summarized and compiled various interferometric space mission proposals (Ni 2016; Table 1). Among the proposed science orbits, there are basically three categories — ASTROD-GW-like, LISA-like and OMEGA-like.

ASTROD-GW-like orbit configuration is to have 3 S/C orbit near two-body L3, L4 and L5 points respectively to form a nearly equilateral triangle: ASTROD-GW (arm length 260 Gm; Ni 2009b, 2009c, 2010, 2012, 2013) having 3 S/C to orbit near Sun-Earth L3, L4 and L5 points; Super-ASTROD (arm length 1300 Gm; Ni 2009a) having 3 S/C to orbit near Sun-Jupiter L3, L4 and L5 points; LAGRANGE (arm length 0.66 Gm, Conklin *et al* 2011) and ASTROD-EM (arm length 0.66 Gm; referred to in Kuroda *et al* 2015 and Ni 2016) having 3 S/C to orbit near Earth-Moon L3, L4 and L5 points.

OMEGA-like science orbits are Earth orbits away from (either inside or outside) Moon's orbit around the Earth. The OMEGA mission proposal (Hiscock and Hellings 1997, Hellings *et al* 2011) consists of six identical spacecraft in a 600,000-km high Earth orbit, two spacecraft at each vertex of a nearly equilateral triangle formation with arm length 1 Gm. OMEGA configuration is outside of Moon's orbit. Mission proposals in this category also include gLISA/GEOGRAWI (arm length 0.073 Gm; Tinto *et al* 2011, 2013, 2015), GADFLI (arm length 0.073 Gm; McWilliams 2011), and TIANQIN (arm length 0.17 Gm; Luo *et al* 2016); these mission configurations are inside Moon's orbit.

LISA—Laser Interferometric Space Antenna—was a proposed ESA-NASA mission which would use coherent laser beams exchanged between three identical spacecraft (S/C) forming a nearly equilateral triangle of side 5 Gm inclined by about 60° with respect to the ecliptic to observe and detect low-frequency cosmic GW (LISA Study Team 2000) (figure 2). The three S/C were designed to be drag-free and to trail the Earth by about 20° in an orbit around the Sun with periods about one year. The formation rotates once per year clockwise or counterclockwise facing the Sun. This project nominally ended with NASA's withdrawal in April, 2011.

After the termination of ESA-NASA collaboration, a down-scaled LISA mission called eLISA/NGO was proposed from a joint effort of seven European countries (France, Germany, Italy, The Netherlands, Spain, Switzerland, and UK) and ESA. The NGO assessment study report received excellent scientific evaluation (http://eLISA-ngo.org). The mission configuration consists of a "mother" S/C at one vertex and two "daughter" S/C at two other vertices with the mother S/C optically linked with two daughter S/C forming an interferometer. The duration of the mission is 2 years for science orbit and about 4 years including transferring and commissioning. The mission S/C orbit configuration is similar to LISA, but with nominal arm length of 1 Gm, inclined by about 60° with respect to the ecliptic, and trailing Earth by 10-20°.

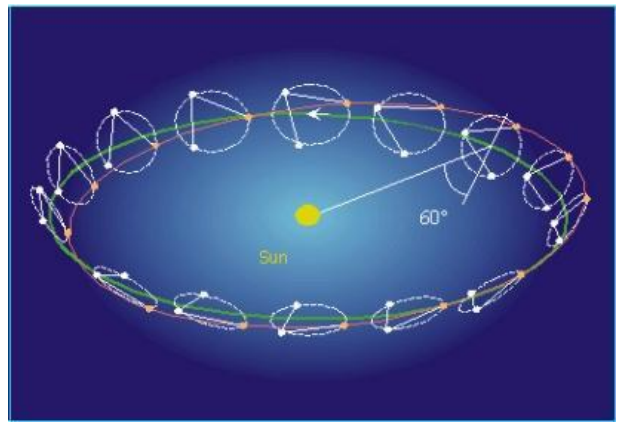

**Figure 2.** Schematic of LISA-type orbit configuration in Earth-like solar orbit. (LISA Study Team 2000)

In a comparative study of TDI's for eLISA/NGO, LISA and ASTROD-GW, we included also an NGO-LISA-type mission with a nominal arm length of 2 Gm (Wang and Ni 2013a).

ALIA (Bender 2004) and TAIJI (Gong *et al* 2015; Wu 2017) have this kind of LISA-like science orbits. The ultimate configuration of Big Bang Observer (Crowder and Cornish 2005) and DECIGO (Kawamura *et al* 2006, 2007, Ando and the DECIGO working group 2013) has 12 spacecraft distributed in the Earth orbit in three groups separated by 120° in orbit; two groups has three spacecraft each in a LISA-like triangular formation and the third group has six spacecraft with two LISA-like triangles forming a star configuration. An alternate configuration is that each group has four spacecraft forming a nearly square configuration (also has a tilt of 60° with respect to the ecliptic plane).

*1.4. New LISA*

A new LISA proposal was recently submitted to ESA on January 13th in response to the call for missions for the L3 slot in the Cosmic Vision Programme (Amaro-Seoane *et al* 2017). On 20 June 2017, ESA announced that "The LISA trio of satellites to detect gravitational waves from space has been selected as the third large-class mission in ESA's Science programme (ESA 2017)." The basic concept is the same as the original LISA with arm length down-scaled to 2.5 Gm. Quoting from the proposal: "The observatory will be based on three arms with six active laser links, between three identical spacecraft in a triangular formation separated by 2.5 million km. Continuously operating heterodyne laser interferometers measure with pm $Hz^{-1/2}$ sensitivity in both directions along each arm, using well-stabilized lasers at 1064 nm delivering 2 W of power to the optical system. Using technology proven in LISA Pathfinder, the Interferometry Measurement System is using optical benches in each spacecraft constructed from an ultra-low expansion glass-ceramic to minimize optical path length changes due to temperature fluctuations. 30 cm telescopes transmit and receive the laser light to and from the other spacecraft. Three independent interferometric combinations of the light travel time between the test masses are possible, allowing, in data processing on the ground, the synthesis of two virtual Michelson interferometers plus a third null-stream, or "Sagnac" configuration." These two virtual Michelson interferometers are two TDIs. They could be two out of three TDI configurations X, Y and Z if they satisfy the noise requirement.

*1.5. TAIJI and other LISA-like missions of various arm lengths*

A feasibility study commissioned by the Chinese Academy of Sciences to explore various possible mission options to detect GWs in space alternative to that of the eLISA/LISA mission concept suggested a few representative mission options descoped from the ALIA mission. The study indicates that, by choosing the arm length of the interferometer to be 3 Gm and shifting the sensitivity floor to around one-hundredth Hz, together with a very moderate improvement on the position noise budget, there are certain mission options capable of exploring light seed, intermediate mass black hole binaries at high redshift that are not readily accessible to eLISA/LISA, and yet the technological requirements are within reach. The feasibility culminated in the TAIJI GW mission concept with 3 Gm arm length LISA-like configuration with more stringent position noise budget (Gong *et al*, 2015, Wu 2017).

For systematic comparison of LISA-like configurations of different arm lengths. We also consider possible mission proposals of nominal arm lengths of 4 Gm and 6 Gm,

*1.6. Comparison of TDIs for interferometers with different arm lengths*

In this paper, we use the CGC 2.7.1 ephemeris (See Appendix) to optimize the orbits and numerically evaluate TDIs. The differences in orbit evolution of Earth calculated using CGC 2.7.1 compared with that of DE430 starting at March 22nd, 2028 for 2200 days are less than 82 m, 0.32 mas and 0.11 mas for radial distance, longitude and latitude respectively.

In table 1, we compile and compare the resulting differences for the first-generation TDIs listed in the subsection *1.2.1.*, i.e. X, Y, Z, X+Y+Z and Sagnac configurations of different arm lengths for various mission proposals -- 1 Gm (eLISA/NGO), 2 Gm (an NGO-LISA-type mission with this nominal arm length), 2.5 Gm (new LISA), 3 Gm (TAIJI), 4 Gm, 5 Gm (original LISA), 6 Gm, and 260 Gm (ASTROD-GW). In table 2, we compile and compare those for second-generation TDIs listed in subsection *1.2.2*.

**Table 1.** Comparison of the resulting path length differences for the first-generation TDI's listed in subsection 1.2.1 (i.e., X, Y, Z, X+Y+Z, Sagnac, U, V, W, P, Q, R, E, F, and G TDI configurations) for various mission proposals with different arm lengths: 1 Gm (eLISA/NGO), 2 Gm (an NGO-LISA-type mission with this nominal arm length), 2.5 Gm (new LISA), 3 Gm (TAIJI), 4 Gm, 5 Gm (original LISA), 6 Gm, and 260 Gm (ASTROD-GW).

| 1st generation TDI configuration | TDI path difference $\Delta L$ | | | | | | | |
|---|---|---|---|---|---|---|---|---|
| | eLISA/NGO [ns] [min, max], rms average | NGO-LISA-type with 2 Gm arm length [ns] [min, max], rms average | New LISA [ns] [min, max], rms average | TAIJI [ns] [min, max], rms average | LISA-type with 4 Gm arm length [ns] [min, max], rms average | LISA-type with 5 Gm arm length [ns] [min, max], rms average | LISA-type with 6 Gm arm length [ns] [min, max], rms average | ASTROD-GW [μs] [min, max], rms average (Wang and Ni, 2015) |
| X | [-119, 99], 40 | [-487, 454], 184 | [-799, 737], 306 | [-1221, 1084], 467 | [-2020, 1748], 812 | [-3286, 2967], 1417 | [-5032, 4466], 2285 | [-194, 182], 111 |
| Y | [-84, 100], 38 | [-434, 359], 176 | [-708, 537], 289 | [-1039, 740], 431 | [-2009, 1837], 917 | [-3543, 3262], 1680 | [-5438, 5201], 2686 | [-190, 196], 113 |
| Z | [-90, 106], 37 | [-339, 405], 152 | [-558, 665], 248 | [-887, 1030], 380 | [-2156, 1894], 826 | [-3724, 3096], 1471 | [-5985, 4902], 2434 | [-200, 194], 115 |
| X+Y+Z | [-0.074, 0.777], 0.243 | [-0.826, 2.910], 0.953 | [-1.754, 4.466], 1.410 | [-3.273, 6.243], 1.949 | [-6.555, 7.909], 3.564 | [-15.693, 8.743], 8.757 | [-31.632, 7.665], 19.397 | [-58, 24], 23 |
| Sagnac-α | [-1965, -1857], 1906, 20* | [-7865, -7401], 7623, 92* | [-12309, -11551], 11911, 153* | [-17759, -16623], 17151, 234* | [-31490, -29650], 30494, 409* | [-49273, -46213], 47644, 715* | [-71096, -66426], 68601, 1155* | [-257627, -257438], 257531, 55* |
| Sagnac-β | [-1948, -1855], 1907, 19* | [-7838, -7434], 7626, 88* | [-12262, -11624], 11915, 144* | [-17666, -16749], 17156, 216* | [-31512, -29626], 30496, 460* | [-49466, -46101], 47649, 843* | [-71437, -66137], 68602, 1349* | [-257626, -257430], 257529, 57* |
| Sagnac-γ | [-1952, -1853], 1906, 18* | [-7799, -7428], 7620, 76* | [-12199, -11593], 11906, 125* | [-17611, -16661], 17145, 192* | [-31610, -29581], 30474, 418* | [-49581, -46142], 47610, 746* | [-71723, -66184], 68546, 1236* | [-257629, -257433], 257530, 57* |
| Relay-U | [-79, 77], 32 | [-298, 297], 137 | [-502, 467], 222 | [-781, 657], 333 | [-1978, 1331], 773 | [-3479, 2322], 1412 | [-5584, 3511], 2295 | [-175, 168], 100 |
| Relay-V | [-102, 80], 34 | [-415, 348], 144 | [-689, 545], 238 | [-1069, 776], 367 | [-1895, 1480], 678 | [-3128, 2507], 1175 | [-4908, 3973], 1940 | [-156, 172], 96 |
| Relay-W | [-79, 96], 35 | [-405, 375], 164 | [-666, 596], 271 | [-987, 879], 407 | [-1809, 1265], 762 | [-3167, 1933], 1370 | [-4865, 2941], 2176 | [-176, 167], 98 |
| Beacon-P | [-60, 49], 20 | [-245, 226], 92 | [-401, 368], 153 | [-612, 541], 233 | [-1010, 876], 406 | [-1639, 1490], 708 | [-2505, 2247], 1141 | [-97, 91], 56 |
| Beacon-Q | [-43, 50], 19 | [-218, 178], 89 | [-356, 267], 145 | [-521, 367], 216 | [-1005, 921], 459 | [-1770, 1637], 841 | [-2711, 2613], 1344 | [-95, 98], 57 |
| Beacon-R | [-46, 53], 19 | [-171, 202], 76 | [-278, 332], 124 | [-442, 514], 190 | [-1076, 948], 413 | [-1856, 1552], 736 | [-2981, 2462], 1218 | [-100, 97], 58 |
| Monitor-E | [-49, 60], 20 | [-226, 245], 92 | [-368, 401], 153 | [-541, 612], 233 | [-876, 1010], 406 | [-1490, 1639], 708 | [-2247, 2505], 1141 | [-91, 97], 56 |
| Monitor-F | [-50, 43], 19 | [-178, 218], 89 | [-267, 356], 145 | [-367, 521], 216 | [-921, 1005], 459 | [-1637, 1770], 841 | [-2613, 2711], 1344 | [-98, 95], 57 |
| Monitor-G | [-53, 46], 19 | [-202, 171], 76 | [-332, 278], 124 | [-514, 442], 190 | [-948, 1076], 413 | [-1552, 1856], 736 | [-2462, 2981], 1218 | [-97, 100], 58 |
| Nominal arm length | 1 Gm (1 Mkm) | 2 Gm | 2.5 Gm | 3 Gm | 4 Gm | 5 Gm | 6 Gm | 260 Gm |
| Mission duration | 2200 days | 2200 days | 2200 days | 2200 days | 2200 days | 2200 days | 2200 days | 10 years |
| Requirement on $\Delta L$ | 10 m (33 ns) | 20 m (67 ns) | 25 m (83 ns) | 30 m (100 ns) | 40 m (133 ns) | 50 m (167 ns) | 60 m (200 ns) | 500 m (1670 ns) |

*root mean square deviation from the mean

**Table 2.** Comparison of the resulting path length differences for the second generation TDIs listed in subsection *1.2.2* (2 arms, *n* = 1: [*ab*, *ba*] (= *abba – baab*), [*cd*, *dc*], [*ef*, *fe*]; 2 arms, *n* = 2: [$a^2b^2$, $b^2a^2$], [*abab*, *baba*], [$ab^2a$, $ba^2b$], [$c^2d^2$, $d^2c^2$], [*cdcd*, *dcdc*], [$cd^2c$, $dc^2d$], [$e^2f^2$, $f^2e^2$], [*efef*, *fefe*], [$ef^2e$, $fe^2f$]; 3 arms: Sagnac-$\alpha 2$, Sagnac-$\beta 2$, and Sagnac-$\gamma 2$,) from different arm lengths for various mission proposals: 1 Gm (eLISA/NGO), 2 Gm (an NGO-LISA-type mission with this nominal arm length), 2.5 Gm (new LISA), 3 Gm (TAIJI), 4 Gm, 5 Gm (original LISA), 6 Gm, and 260 Gm (ASTROD-GW).

| 2nd Generation TDI configuration | TDI path difference ΔL | | | | | | | |
|---|---|---|---|---|---|---|---|---|
| | eLISA/ NGO [ns] [min, max], rms average | NGO-LISA-type with 2 Gm arm length [ns] [min, max], rms average | New LISA [ns] [min, max], rms average | TAIJI [ns] [min, max], rms average | LISA-type with 4 Gm arm length [ns] [min, max], rms average | LISA-type with 5 Gm arm length [ns] [min, max], rms average | LISA-type with 6 Gm arm length [ns] [min, max], rms average | ASTROD-GW [µs] [min, max], rms average (Wang and Ni, 2015) |
| [*ab*, *ba*] | [-0.51, 0.45], 0.187 | [-4.4, 3.3], 1.6 | [-8.9, 6.5], 3.2 | [-15.8, 11.6], 5.7 | [-35.3, 28.5], 13.5 | [-72, 60], 28 | [-130, 113], 51 | [-252, 244], 152 |
| [*cd*, *dc*] | [-0.44, 0.45], 0.179 | [-3.6, 3.8], 1.6 | [-7.3, 7.4], 3.1 | [-12.8, 12.6], 5.4 | [-35.4, 26.6], 13.9 | [-76, 53], 30 | [-140, 89], 55 | [-248, 258], 156 |
| [*ef*, *fe*] | [-0.44, 0.48], 0.182 | [-3.5, 3.8], 1.5 | [-7.1, 7.3], 3.0 | [-13.0, 12.6], 5.2 | [-35.0, 30.0], 13.3 | [-73, 62], 28 | [-135, 111], 52 | [-255, 256], 155 |
| [$a^2b^2$, $b^2a^2$] | [-4.1, 3.6], 1.5 | [-35, 26], 13 | [-71, 52], 26 | [-126, 93], 45 | [-282, 228], 107 | [-576, 479], 222 | [-1033, 898], 404 | [-2011, 1949], 1209 |
| [$c^2d^2$, $d^2c^2$] | [-3.5, 3.5], 1.4 | [-29, 30], 12 | [-58, 59], 24 | [-102, 101], 43 | [-283, 213], 111 | [-607, 421], 237 | [-1115, 712], 435 | [-1977, 2063], 1248 |
| [$e^2f^2$, $f^2e^2$], | [-3.4, 3.8], 1.5 | [-28, 30], 12 | [-56, 58], 23 | [-103, 101], 41 | [-279, 239], 106 | [-578, 494], 221 | [-1075, 881], 409 | [-2038, 2043], 1240 |
| [*abab*, *baba*] | [-2.1, 1.8], 0.8 | [-18, 13], 6 | [-36, 26], 13 | [-63, 46], 23 | [-141, 114], 54 | [-288, 239], 111 | [-517, 449], 202 | [-1006, 975], 605 |
| [*cdcd*, *dcdc*] | [-1.7, 1.8], 0.7 | [-14, 15], 6 | [-29, 30], 12 | [-51, 50], 21 | [-141, 106], 55 | [-304, 210], 118 | [-557, 356], 218 | [-989, 1032], 624 |
| [*efef*, *fefe*] | [-1.7, 1.9], 0.7 | [-14, 15], 6 | [-28, 29], 12 | [-52, 50], 21 | [-140, 120], 53 | [-289, 247], 111 | [-538, 440], 204 | [-1019, 1022], 620 |
| [$ab^2a$, $ba^2b$] | [-0.01, 0.01], 0.002 | [-0.02, 0.01], 0.002 | [-0.02, 0.01], 0.003 | [-0.02, 0.01], 0.003 | [-0.02, 0.01], 0.004 | [-0.02, 0.02], 0.005 | [-0.03, 0.02], 0.007 | [-0.91, 9.99], 0.50 |
| [$cd^2c$, $dc^2d$] | [-0.011, 0.010], 0.002 | [-0.018, 0.011], 0.002 | [-0.012, 0.009], 0.002 | [-0.011, 0.010], 0.002 | [-0.016, 0.012], 0.003 | [-0.021, 0.014], 0.004 | [-0.027, 0.019], 0.006 | [-0.69, 8.02], 0.50 |
| [$ef^2e$, $fe^2f$] | [-0.010, 0.013], 0.002 | [-0.013, 0.012], 0.002 | [-0.013, 0.009], 0.002 | [-0.017, 0.012], 0.002 | [-0.017, 0.012], 0.003 | [-0.022, 0.014], 0.005 | [-0.027, 0.018], 0.007 | [-19.39, 1.04], 0.60 |
| Sagnac-$\alpha 2$ | [-0.17, 0.18], 0.07 | [-1.3, 1.5], 0.7 | [-2.7, 3.0], 1.3 | [-4.8, 5.2], 2.2 | [-11.6, 11.5], 5.4 | [-23.2, 22.8], 11.2 | [-42.0, 39.8], 20.6 | [-92, 97], 57 |
| Sagnac-$\beta 2$ | [-0.17, 0.16], 0.07 | [-1.5, 1.3], 0.6 | [-3.0, 2.6], 1.2 | [-5.2, 4.4], 2.1 | [-11.1, 11.8], 5.3 | [-22.7, 24.6], 11.3 | [-39.9, 44.2], 21.0 | [-94, 92], 58 |
| Sagnac-$\gamma 2$ | [-0.19, 0.16], 0.07 | [-1.5, 1.4], 0.6 | [-2.9, 2.7], 1.2 | [-5.0, 4.9], 2.1 | [-12.3, 12.2], 5.3 | [-26.3, 25.4], 11.1 | [-48.3, 46.9], 20.8 | [-92, 96], 58 |
| Nominal arm length | 1 Gm (1 Mkm) | 2 Gm | 2.5 Gm | 3 Gm | 4 Gm | 5 Gm | 6 Gm | 260 Gm |
| Mission duration | 2200 days | 2200 days | 2200 days | 2200 days | 2200 days | 2200 days | 2200 days | 10 years |
| Requirement on ΔL | 10 m (33 ns) | 20 m (67 ns) | 25 m (83 ns) | 30 m (100 ns) | 40 m (133 ns) | 50 m (167 ns) | 60 m (200 ns) | 500 m (1670 ns) |

From table 1, all the first generation TDIs for LISA-like missions do not satisfy their respective requirements. From table 2, all the second generation TDIs for LISA-like missions satisfy their respective requirements with good margins. To use first-generation TDIs, requirement must be relaxed with accompanying technology development. To use second-generation TDIs, the corresponding GW response and sensitivity must be calculated.

In section 2, we work out a set of 2200-day optimized spacecraft orbits starting at March 22, 2028 using the CGC 2.7.1 ephemeris framework for new LISA, TAIJI, and other LISA-like missions with arm lengths 1, 2, 4, 5, 6 Gm. In section 3, we obtain the numerical results pertaining to the first-generation TDIs listed in table 1. In section 4, we obtain the numerical results pertaining to the second-generation TDIs listed in table 2. In section 5, we compare and discuss in detail the resulting differences due to different arm lengths for various mission proposals including ASTROD-GW, and conclude this paper with discussion and outlook.

## 2. New LISA, TAIJI and other LISA-like mission orbit optimizations

In the LISA-like missions, the distance of any two of three spacecraft must be maintained as close as possible during geodetic flight to minimize relative Doppler velocities between spacecraft for satisfying respective Doppler frequency requirements. LISA orbit configuration has been studied analytically and numerically in various previous works (Vincent and Bender, 1987; Folkner *et al*, 1997; Cutler, 1998; Hughes, 2002; Hechler and Folkner, 2003; Dhurandhar *et al*, 2005; Yi *et al*, 2008; Li *et al*, 2008; Dhurandhar, Ni, and Wang 2013). We have also used the CGC ephemeris framework to numerically optimize the orbit configuration of eLISA/NGO (Wang and Ni, 2013a) and a LISA-type mission with 2 · Gm nominal arm length (Wang and Ni, 2013a), as well as that of ASTROD-GW (Men *et al*, 2009, 2010; Wang and Ni, 2011, 2012, 2013b).

The proposal of new LISA (Amaro-Seoane *et al*, 2017) chose as its nominal arm length 2.5 Gm. TAIJI GW mission proposal chose as its nominal arm length 3 Gm. For comparison, we also work out in this paper orbit designs and TDIs in the same mission period as new LISA for LISA-type missions with 1 Gm (eLISA/NGO), 2 Gm, 4 Gm, 5 Gm (original LISA), and 6 Gm nominal arm lengths.

In this section, we describe the procedures of orbit choice and work out the optimization by taking the examples of the new LISA proposal with nominal arm length equal to 2.5 Gm and the TAIJI proposal with arm length targeting at 3.0 Gm, respectively. After that, we work out the results for optimized orbits of other LISA-like mission proposals.

### *2.1. The initial choice of initial conditions for the new LISA*

There are various ways to choose the orbits of the three spacecraft so that the orbit configuration satisfying the equal arm length requirement to first order in $\alpha$ [$= l/(2R)$], the ratio of the planned arm length $l$ of the orbit configuration to twice radius $R$ (1 AU) of the mean Earth orbit. We follow the procedures given in Dhurandhar *et al* (2005) and our previous paper (Wang and Ni, 2013a) to make initial choice of initial conditions to start our optimitization procedure.

Choosing the initial time $t_0$ for science orbit configuration to be JD2462503.0 (2030-Jan-1st 12:00:00), we work in the Heliocentric Coordinate System ($X$, $Y$, $Z$). $X$-axis is in the direction of vernal equinox. First, as in Dhurandhar *et al* (2005), a set of elliptical S/C orbits is defined as

$$X_f = R(\cos\psi_f + e)\cos\varepsilon,$$
$$Y_f = R(1-e^2)^{1/2}\sin\psi_f,$$
$$Z_f = R(\cos\psi_f + e)\sin\varepsilon, \qquad (2.1)$$

where for the mission with nominal arm length $l$ equal $\lambda$ Gm, $e = 0.001925 \times \lambda$; $\varepsilon = 0.00333 \times \lambda$; $R = 1$ AU.. The eccentric anomaly $\psi_f$ is related to the mean anomaly $\Omega(t - t_0)$ by

$$\psi_f + e\sin\psi_f = \Omega(t-t_0), \qquad (2.2)$$

where $\Omega$ is defined as $2\pi/$(one sidereal year). The eccentric anomaly $\psi_f$ can be solved by numerical iteration. Define $\psi_k$ to be implicitly given by

$$\psi_k + e\sin\psi_k = \Omega(t-t_0) - 120°(k-1), \text{ for } k=1,2,3. \qquad (2.3)$$

Define $X_{fk}$, $Y_{fk}$, $Z_{fk}$, ($k$ = 1.2, 3) to be

$$X_{fk} = R(\cos\psi_k + e)\cos\varepsilon,$$
$$Y_{fk} = R(1-e^2)^{1/2}\sin\psi_k,$$
$$Z_{fk} = R(\cos\psi_k + e)\sin\varepsilon. \qquad (2.4)$$

Define $X_{f(k)}$, $Y_{f(k)}$, $Z_{f(k)}$, ($k = 1, 2, 3$), i.e., $X_{f(1)}$, $Y_{f(1)}$, $Z_{f(1)}$; $X_{f(2)}$, $Y_{f(2)}$, $Z_{f(2)}$; $X_{f(3)}$, $Y_{f(3)}$, $Z_{f(3)}$ to be

$$X_{f(k)} = X_{fk} \cos[120°(k-1)+\varphi_0] - Y_{fk} \sin[120°(k-1)+\varphi_0],$$
$$Y_{f(k)} = X_{fk} \sin[120°(k-1)+\varphi_0] + Y_{fk} \cos[120°(k-1)+\varphi_0],$$
$$Z_{f(k)} = Z_{fk}.$$
(2.5)

Here $\varphi_0 \equiv \psi_E - 20°$ and $\psi_E$ is defined to be the position angle of Earth w.r.t. the $X$-axis at $t_0$. The three S/C orbits are (for one-body central problem) are

$$\mathbf{R}_{S/C1} = (X_{f(1)}, Y_{f(1)}, Z_{f(1)}),$$
$$\mathbf{R}_{S/C2} = (X_{f(2)}, Y_{f(2)}, Z_{f(2)}),$$
$$\mathbf{R}_{S/C3} = (X_{f(3)}, Y_{f(3)}, Z_{f(3)}).$$
(2.6)

The initial positions can be obtained by choosing $t = t_0$ and the initial velocities by calculating the derivatives w.r.t. time at $t = t_0$. With the choice of $t_0 =$ JD2462503.0 (2030-Jan-1st 12:00:00), the initially chosen orbits have relatively good equal-arm performance until JD2464053.0 (2034-Mar-31st 12:00:00) From the time trend of the performance we perceived that the orbits would still be rather good when we back evolved the orbit for a period of time; it is indeed so back to JD2461853.0 (2028-Mar-22nd 12:00:00). Thereby, there could be a promising duration of 2200 days for optimization when the mission is set to start from JD2461853.0 (2028-Mar-22nd 12:00:00) with the evolved back initial conditions. According to the new LISA proposal (Amaro-Seoane *et al*, 2017), all the technology would be demonstrated at least to level 6 before 2020, therefore, pending on budget allocation, launch in 2027 or late 2026 would feasible and GW observation starting in 2028 would be a possible scenario.

The goal of the orbit optimization is to equalize the three arm lengths of the mission formation and to reduce the relative line-of-sight velocities (Doppler velocities) between three pairs of spacecraft as much as possible. For the new LISA proposal of 2.5 Gm arm length, the requirement on the Doppler velocities is below ±5 m/s between the S/Cs in order for frequency tracking between spacecraft to be within ±5 MHz (for laser light of 1064 nm wavelength) due to Doppler frequency shifts (Amaro-Seoane *et al* 2017). For TDI, minimize the Doppler velocities between the S/C also minimize the path length differences of various TDI configurations. We assume that the requirement on the Doppler velocities is directly proportional to the arm length. For example, for the original LISA with arm length 5 Gm the requirement is ±10 m/s (LISA Study Team, 2000), for TAIJI of 3.0 Gm arm length the Doppler velocities is required to be smaller than 6 m/s, and by that analogy, 12 m/s should be upper bound for 6.0 Gm arm length. To accelerate our optimization program, Runge-Kutta 7th/8th order integration is used to search for the orbit in accordance with the mission requirement. The 4th order Runge-Kutta is used to verify stability after one candidate orbit has been achieved.

*2.2. The new LISA mission orbit optimization*

The goal of the New LISA mission orbit optimization is to equalize the three arm lengths of the new LISA formation and to make the relative line-of-sight velocities smaller than 5 m/s between three pairs of spacecraft. Firstly, the initial conditions for the 3 S/Cs are calculated from the equations in section 2.1 at JD2462503.0 (2030-Jan-1st 12:00:00) and evolved back to JD2461853.0 (2028-Mar-22nd 12:00:00) using CGC 2.7.1 ephemeris framework. We list this initial choice of initial states in the third column of table 3. The LISA spacecraft orbits are then calculated for 2200 days using CGC 2.7.1. The variations of arm lengths and Doppler velocities between the LISA S/Cs are drawn in figure 3. We note that science observation period for new LISA has not been determined. However, the expected technical readiness will exceed level 6 by 2020 and the optimistic observation start for new LISA could be in 2028. For other starting dates, similar orbits could be worked out.

As we can see in the figure 3, the Doppler velocity between S/C1 and S/C3 goes slightly beyond 5 m/s. The optimization procedure is to modify the initial heliocentric distances and/or velocities of the S/C.

So we tentatively adjust the initial heliocentric distance of S/C1 to optimize the orbit. One orbit meet the mission requirement achieved when heliocentric distance of S/C1 is decreased by a factor of $3.0 \times 10^{-7}$. The initial condition of optimized orbit is listed in the fifth column of table 3. Figure 4 shows the variations of the arm lengths, Doppler velocities, the formation angles, and the lag angle behind the Earth with the mission time using these initial conditions. The variations of arm lengths (within ±1%) and velocities in the line of sight direction (within ±5.0 m/s requirement) are achieved. The angle between barycentre of S/C and Earth in 2200 days starts at 22° behind Earth and varies between 18° and 23° with a quasi-period of variation about 1 sidereal year due mainly to Earth's elliptic motion.

The new LISA proposes the configuration lag angle behind the Earth to be around 20° while eLISA/NGO proposes the lag angle to be around 10°. This may make the orbits of the New LISA S/Cs suffer less gravitational perturbation from the Earth and the orbit configuration can stay stable for more than 6 years comparing to the orbit configuration of eLISA/NGO (Wang and Ni, 2013a).

**Table 3.** Initial states (conditions) of 3 S/C of New LISA mission with 2.5 Gm arm length at epoch JD2461853.0 for our initial choice, and after optimizations in J2000 equatorial (Earth mean equator and equinox) solar-system-barycentric coordinate system

|   |   | Initial choice of S/C initial states |   | Initial states of S/C after final optimization |
|---|---|---|---|---|
| S/C1 Position (AU) | X | -9.342358697598E-01 | adjust to ==> | -9.342355891858E-01 |
|  | Y | 3.222028021891E-01 |  | 3.222027047288E-01 |
|  | Z | 1.415510901823E-01 |  | 1.415510473840E-01 |
| S/C1 Velocity (AU/day) | $V_x$ | -6.020533666442E-03 | = | -6.020533666442E-03 |
|  | $V_y$ | -1.471303796371E-02 |  | -1.471303796371E-02 |
|  | $V_z$ | -6.532104563056E-03 |  | -6.532104563056E-03 |
| S/C2 Position (AU) | X | -9.422917194822E-01 | = | -9.422917194822E-01 |
|  | Y | 3.075956329521E-01 |  | 3.075956329521E-01 |
|  | Z | 1.403200701890E-01 |  | 1.403200701890E-01 |
| S/C2 Velocity (AU/day) | $V_x$ | -5.875601408922E-03 | = | -5.875601408922E-03 |
|  | $V_y$ | -1.480936170059E-02 |  | -1.480936170059E-02 |
|  | $V_z$ | -6.319195852807E-03 |  | -6.319195852807E-03 |
| S/C3 Position (AU) | X | -9.335382669969E-01 | = | -9.335382669969E-01 |
|  | Y | 3.132742531958E-01 |  | 3.132742531958E-01 |
|  | Z | 1.273476800288E-01 |  | 1.273476800288E-01 |
| S/C3 Velocity (AU/day) | $V_x$ | -5.949351791423E-03 | = | -5.949351791423E-03 |
|  | $V_y$ | -1.490443611747E-02 |  | -1.490443611747E-02 |
|  | $V_z$ | -6.410590762560E-03 |  | -6.410590762560E-03 |

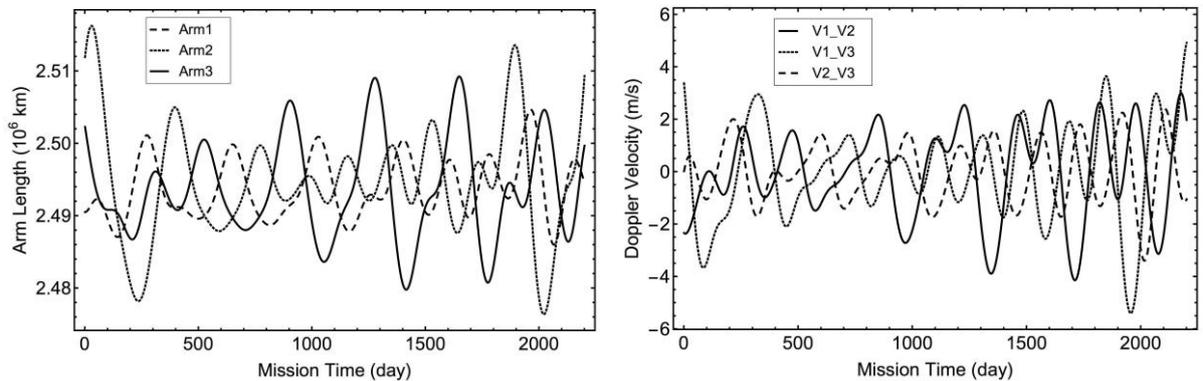

**Figure 3.** Variations of the arm lengths and the velocities in the line of sight direction in 2200 days for the new LISA S/C configuration with initial conditions given in column 3 (initial choice) of table 3.

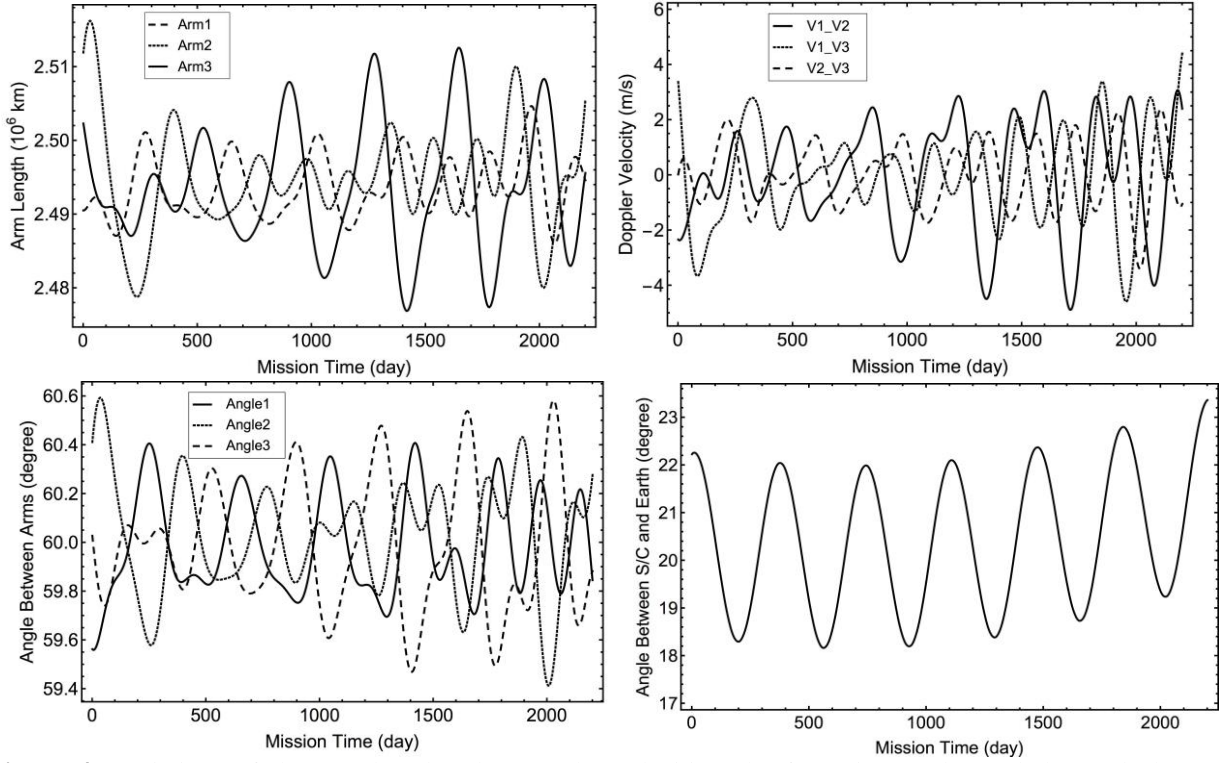

**Figure 4.** Variations of the arm lengths, the Doppler velocities, the formation angles and the angle between barycentre of S/C and Earth in 2200 days for the new LISA S/C configuration with initial conditions given in column 4 (after final optimization) of table 3.

*2.3. The initial choice of initial conditions for TAIJI and its orbit optimization*

TAIJI proposes a LISA-like configuration with 3.0 Gm nominal arm length for GW detection. The procedure for initial orbit choice and optimization is same as new LISA. The initial conditions for the 3 S/Cs are calculated from the equations in section 2.1 at JD2462503.0 (2030-Jan-1st 12:00:00) and evolved back to JD2461853.0 (2028-Mar-22nd 12:00:00) using CGC 2.7.1 ephemeris framework. We list this initial choice of initial states in the third column of table 4. The TAIJI spacecraft orbits are then calculated for 2200 days using CGC 2.7.1. The variations of arm lengths and Doppler velocities between the TAIJI S/Cs are drawn in figure 5.

Comparing to figure 3, the curves in figure 5 have similar trends in time. Same procedure was applied to the adjustment. The optimized orbit was achieved when the heliocentric distance of S/C1 was decreased by a factor of $2.7 \times 10^{-7}$, and the initial conditions are shown in the fifth column of table 4. The final achieved orbit's variations of the arm lengths, Doppler velocities, the formation angles, and the lag angle behind the Earth with the mission time are shown in figure 6. The variations of arm lengths are within ±1% and velocities in the line of sight direction are within ±6.0 m/s requirement. The angle between barycentre of S/C and Earth in 2200 days starts at 22° behind Earth and varies between 18° and 23° with a quasi-period of variation about 1 sidereal year due mainly to Earth's elliptic motion.

**Table 4.** Initial states (conditions) of 3 S/C of TAIJI at epoch JD2461853.0 for our initial choice (third column) and after optimizations (fifth column) in J2000 equatorial (Earth mean equator and equinox) solar-system-barycentric coordinate system.

|  |  | Initial choice of S/C initial states |  | Initial states of S/C after final optimization |
|---|---|---|---|---|
| S/C1 Position (AU) | X | -9.337345684115E-01 | adjust to ==> | -9.337343160303E-01 |
|  | Y | 3.237549276553E-01 |  | 3.237548395220E-01 |
|  | Z | 1.426066025785E-01 |  | 1.426065637750E-01 |
| S/C1 Velocity (AU/day) | $V_x$ | -6.034814754038E-03 | = | -6.034814754038E-03 |
|  | $V_y$ | -1.469355864558E-02 |  | -1.469355864558E-02 |
|  | $V_z$ | -6.554198841518E-03 |  | -6.554198841518E-03 |

| S/C2 Position (AU) | X | -9.433977273640E-01 | = | -9.433977273640E-01 |
| | Y | 3.062344469040E-01 | | 3.062344469040E-01 |
| | Z | 1.411270887844E-01 | | 1.411270887844E-01 |
| S/C2 Velocity (AU/day) | $V_x$ | -5.861017364349E-03 | = | -5.861017364349E-03 |
| | $V_y$ | -1.480919323217E-02 | | -1.480919323217E-02 |
| | $V_z$ | -6.298978166673E-03 | | -6.298978166673E-03 |
| S/C3 Position (AU) | X | -9.328957809408E-01 | = | -9.328957809408E-01 |
| | Y | 3.130424089270E-01 | | 3.130424089270E-01 |
| | Z | 1.255542247698E-01 | | 1.255542247698E-01 |
| S/C3 Velocity (AU/day) | $V_x$ | -5.949486480991E-03 | = | -5.949486480991E-03 |
| | $V_y$ | -1.492350292755E-02 | | -1.492350292755E-02 |
| | $V_z$ | -6.408454202380E-03 | | -6.408454202380E-03 |

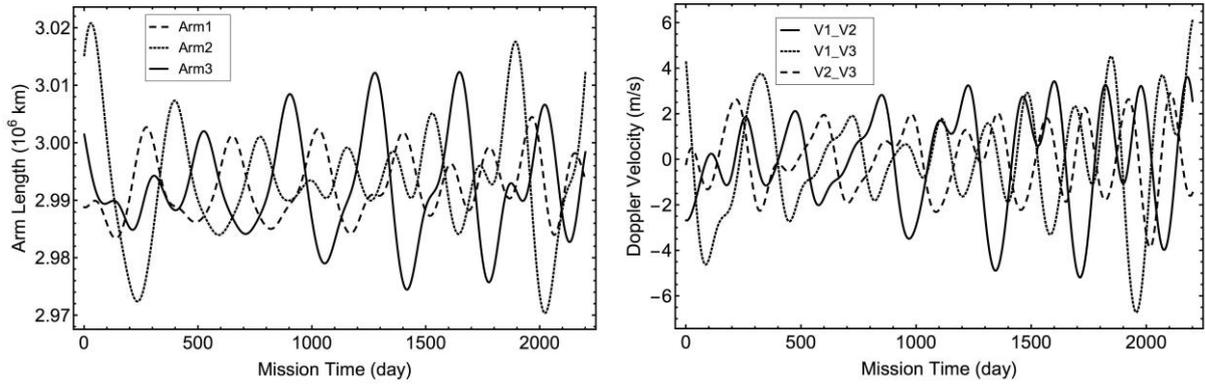

**Figure 5.** Variations of the arm lengths and the velocities in the line of sight direction in 2200 days for the TAIJI S/C configuration with initial conditions given in column 3 (initial choice) of table 4.

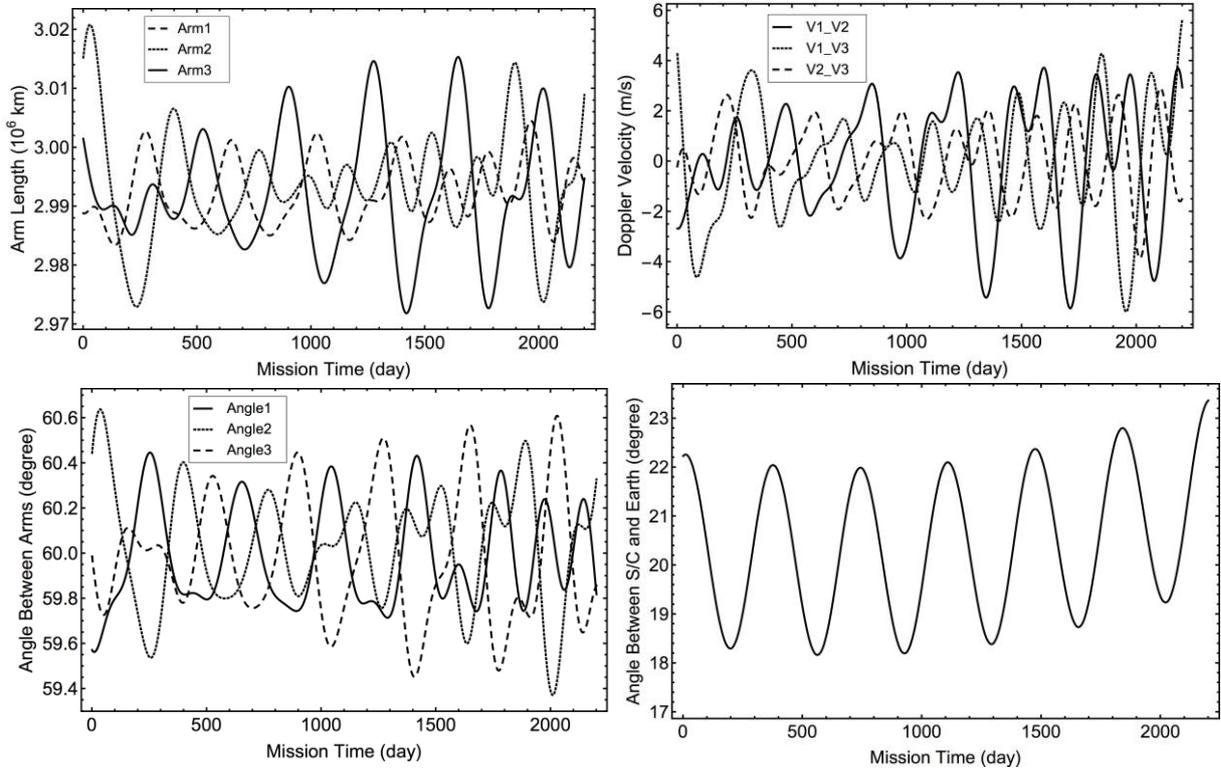

**Figure 6.** Variations of the arm lengths, the velocities in the line of sight direction, the formation angles, (add this figure) and angle between barycentre of S/C and Earth in 2200 days for the TAIJI S/C configuration with initial conditions given in column 5 (after optimization) of table 4.

## 2.4. The initial choice of initial conditions for other LISA-like missions of various arm lengths and their orbit optimization

For LISA-like missions with different am lengths, the goal of the optimization is to make the Doppler velocities under a given prorated requirement. We assume that the requirement increase by 2 m/s per Gm ($10^6$ km) increase on the nominal arm length. In general, the shorter arm length the mission formation, the easier to achieve the optimization (since linear approximation works better). For the 1.0 Gm arm length mission orbits, the initial conditions specified from the equations in subsection *2.1* with back evolution satisfy the requirement that the line-of-sight Doppler velocities be smaller than ± 2 m/s for the period of mission considered. For the cases of the arm lengths 2.0 Gm, 2.5 Gm and 3.0 Gm, the initial conditions from the analytical equation almost satisfy the mission requirement. Only slight adjustment of the initial heliocentric distance of S/C1 is needed.

With the increase of the arm length to 4.0 Gm, 5.0 Gm and 6.0 Gm, there are larger nonlinear perturbations on the variation of the arm length. Therefore more iterative adjustments are needed to meet the mission requirements. For the 4.0 Gm and 5.0 Gm cases, we adjusted the heliocentric distances of S/C1 and S/C2 together with the heliocentric velocity of the S/C1, before we achieve final optimized orbit. The adjustment on S/C3 position was also used when we optimized the orbit for the 6 Gm arm length case. The scale of the adjustment is the order of $10^{-6}$ to $10^{-7}$ in the ecliptic heliocentric coordinate system. The optimized initial conditions for the 5 cases of different arm lengths considered in this subsection are shown in table 5. More information can be found in figures 1-5 the Supplement of this article.

Since the initial condition choice for all missions worked out in this paper share the same epoch and nearly same barycenter of S/Cs, there could be some common features when we optimize this family which would be beneficial to our optimization process. Also we believe that the optimized orbits we achieved in this paper are definitely not the only choices. These solutions just illustrate the possibilities and what we can achieve and assume. From our previous investigations and present investigation, there should be solutions at any epochs.

**Table 5.** Initial states (conditions) of 3 S/C of optimized LISA-type mission with different arm lengths at epoch JD2461853.0 (2028-Mar-22nd 12:00:00) in J2000 equatorial (Earth mean equator and equinox) solar-system-barycentric coordinate system

| Arm Length (Gm) | (AU) /(AU/day) | | S/C1 | S/C2 | S/C3 |
|---|---|---|---|---|---|
| 1 | Position | X | -9.357297393390E-01 | -9.389558953532E-01 | -9.354523832766E-01 |
| | | Y | 3.175249022108E-01 | 3.116740035186E-01 | 3.139517985143E-01 |
| | | Z | 1.384166686585E-01 | 1.379265627712E-01 | 1.327436476580E-01 |
| | Velocity | $V_x$ | -5.977487755471E-03 | -5.919394364688E-03 | -5.948925229945E-03 |
| | | $V_y$ | -1.477105821259E-02 | -1.480954135324E-02 | -1.484735001743E-02 |
| | | $V_z$ | -6.465498086333E-03 | -6.380065664383E-03 | -6.416821125538E-03 |
| 2 | Position | X | -9.347352669162E-01 | -9.411827396003E-01 | -9.341785350444E-01 |
| | | Y | 3.206469987577E-01 | 3.089559597107E-01 | 3.135030989930E-01 |
| | | Z | 1.405008890136E-01 | 1.395176390651E-01 | 1.291437392093E-01 |
| | Velocity | $V_x$ | -6.006218730732E-03 | -5.890192320964E-03 | -5.949213312432E-03 |
| | | $V_y$ | -1.473244784149E-02 | -1.480947601134E-02 | -1.488538837914E-02 |
| | | $V_z$ | -6.509956210443E-03 | -6.339449669282E-03 | -6.412697381390E-03 |
| 4 | Position | X | -9.327266874930E-01 | -9.455998489267E-01 | -9.316041884175E-01 |
| | | Y | 3.268482296518E-01 | 3.035092228176E-01 | 3.125697220925E-01 |
| | | Z | 1.447335875947E-01 | 1.427547210038E-01 | 1.219751478289E-01 |
| | Velocity | $V_x$ | -6.063271925217E-03 | -5.831870261806E-03 | -5.949744254084E-03 |
| | | $V_y$ | -1.465438542511E-02 | -1.480869462463E-02 | -1.496169386072E-02 |
| | | $V_z$ | -6.598221078427E-03 | -6.258651209747E-03 | -6.404090921964E-03 |
| 5 | Position | X | -9.317122569544E-01 | -9.477904576171E-01 | -9.303038243396E-01 |
| | | Y | 3.299269994748E-01 | 3.007808152754E-01 | 3.120850334818E-01 |
| | | Z | 1.468818648738E-01 | 1.444007020429E-01 | 1.184065522341E-01 |
| | | $V_x$ | -6.091593636329E-03 | -5.802751784201E-03 | -5.949986166082E-03 |
| | Velocity | $V_y$ | -1.461494061048E-02 | -1.480798178629E-02 | -1.499996136965E-02 |
| | | $V_z$ | -6.642025921601E-03 | -6.218468839601E-03 | -6.399606859148E-03 |
| 6 | Position | X | -9.306816929936E-01 | -9.499585655003E-01 | -9.289944758125E-01 |
| | | Y | 3.330008645251E-01 | 2.980619470362E-01 | 3.115882451472E-01 |
| | | Z | 1.490558508785E-01 | 1.460701491120E-01 | 1.148484474658E-01 |
| | Velocity | $V_x$ | -6.120039892635E-03 | -5.773977554200E-03 | -5.950211771694E-03 |
| | | $V_y$ | -1.457521032961E-02 | -1.480703342030E-02 | -1.503830563769E-02 |
| | | $V_z$ | -6.685605443277E-03 | -6.178414964039E-03 | -6.395001318691E-03 |

## 3. Numerical simulation of the first-generation TDI for new LISA and TAIJI and other LISA-like missions

In our early papers (Wang and Ni, 2011, 2012; Wang, 2011), we have used the CGC 2.7 ephemeris framework to calculate the difference between the two path lengths for first-generation TDI configurations of the planar non-precession ASTROD-GW orbit configuration. The results were shown by plotting the difference as function of the epoch of ASTROD-GW orbit configuration. The method of obtaining these solutions and the TDI configurations were briefly reviewed in section *1.2*.

In the numerical calculation in this section, we calculate the difference between the two path lengths for TDI configurations and plotted the difference as function of the signal arriving epoch of TDI in first-generation TDI configurations --- Michelson X, Y & Z; Sagnac $\alpha$, $\beta$ & $\gamma$; Relay U, V & W; Beacon P, Q & R; Monitor E, F & G for the new LISA mission and the TAIJI mission together with LISA-like missions of arm length 1 Gm, 2 Gm, 4 Gm, 5 Gm and 6 Gm. We make use of the iteration and interpolation methods (Chiou and Ni, 2000a, 2000b; Newhall, 1989; Li and Tian, 2004) to calculate the time in the barycentric coordinate system as in our early papers. We do this for the new LISA mission in section 3.1, for the TAIJI mission in section 3.2, and for the LISA-like missions of arm length 1 Gm, 2 Gm, 4 Gm, 5 Gm and 6 Gm in section 3.3.

*3.1. Numerical results of the first-generation TDI for new LISA*

*3.1.1. Unequal-arm Michelson* X*, Y & Z TDIs and their sum* X+Y+Z *for new LISA*

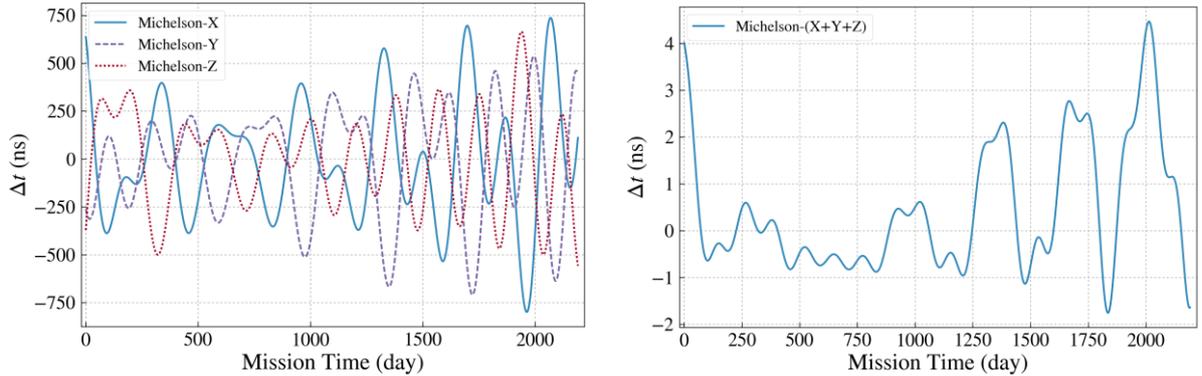

**Figure 7.** The optical path length differences for Unequal-arm Michelson X, Y, & Z TDIs (left panel) and their sum X + Y + Z (right panel). There is a clear cancellation of optical path length differences by 2 orders of magnitudes in the sum.

### 3.1.2. Sagnac α, β & γ; Relay U, V & W; Beacon P, Q & R; Monitor E, F & G TDIs for new LISA

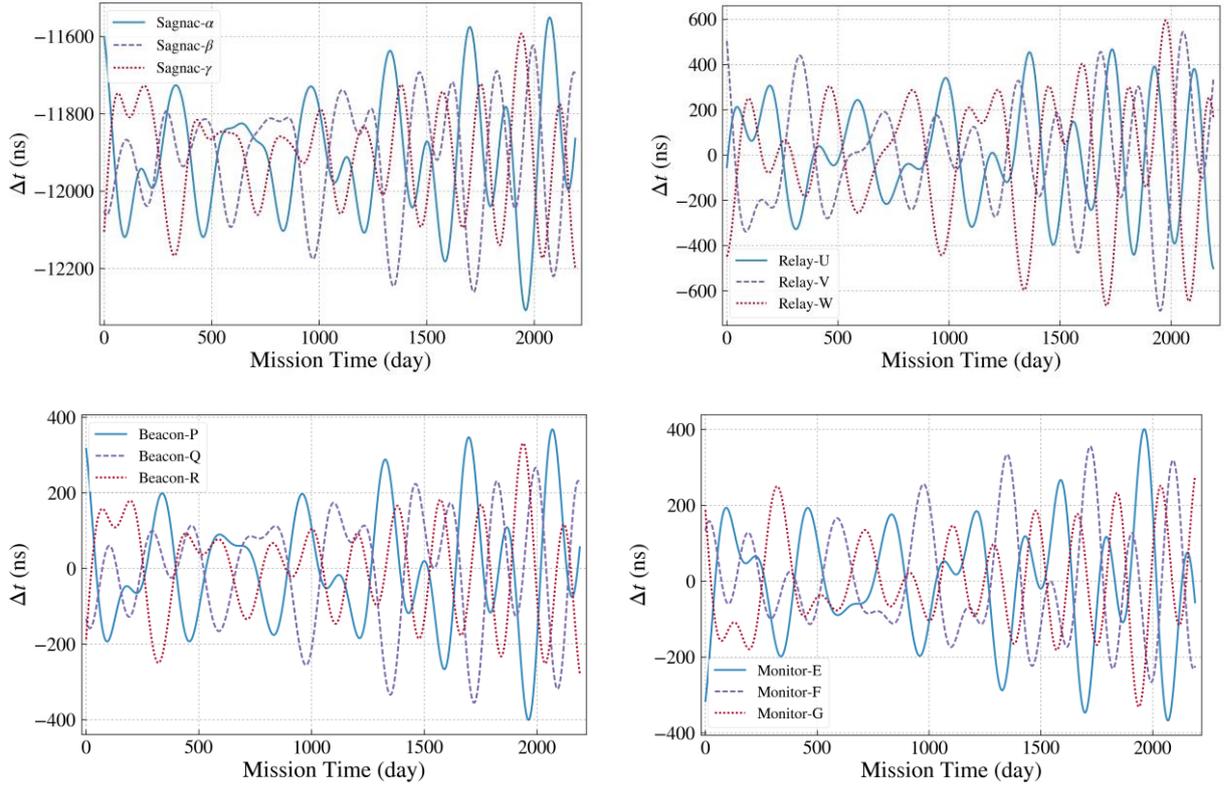

**Figure 8.** The optical path length differences vs time epoch for Sagnac α, β & γ; Relay U, V & W; Beacon P, Q & R; Monitor E, F & G TDIs for new LISA.

## 3.2. Numerical results of the first-generation TDI for TAIJI

### 3.2.1. Unequal-arm Michelson X, Y & Z TDIs and their sum X+Y+Z for TAIJI

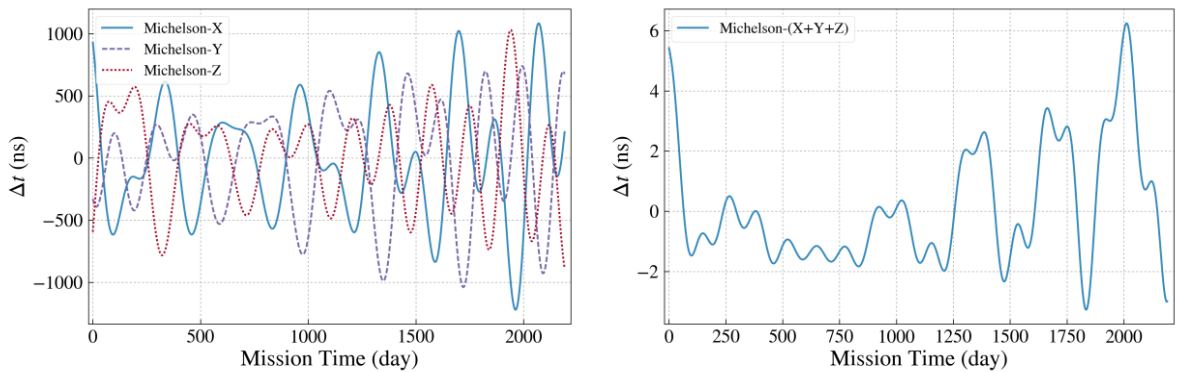

**Figure 9.** The optical path length differences for Unequal-arm Michelson X, Y, & Z TDIs (left) and their sum X + Y + Z (right) for TAIJI. There is a clear cancellation of optical path length differences by 2 orders of magnitudes in the sum.

### 3.2.2. Sagnac α, β & γ; Relay U, V & W; Beacon P, Q & R; Monitor E, F & G first-generation TDIs for TAIJI

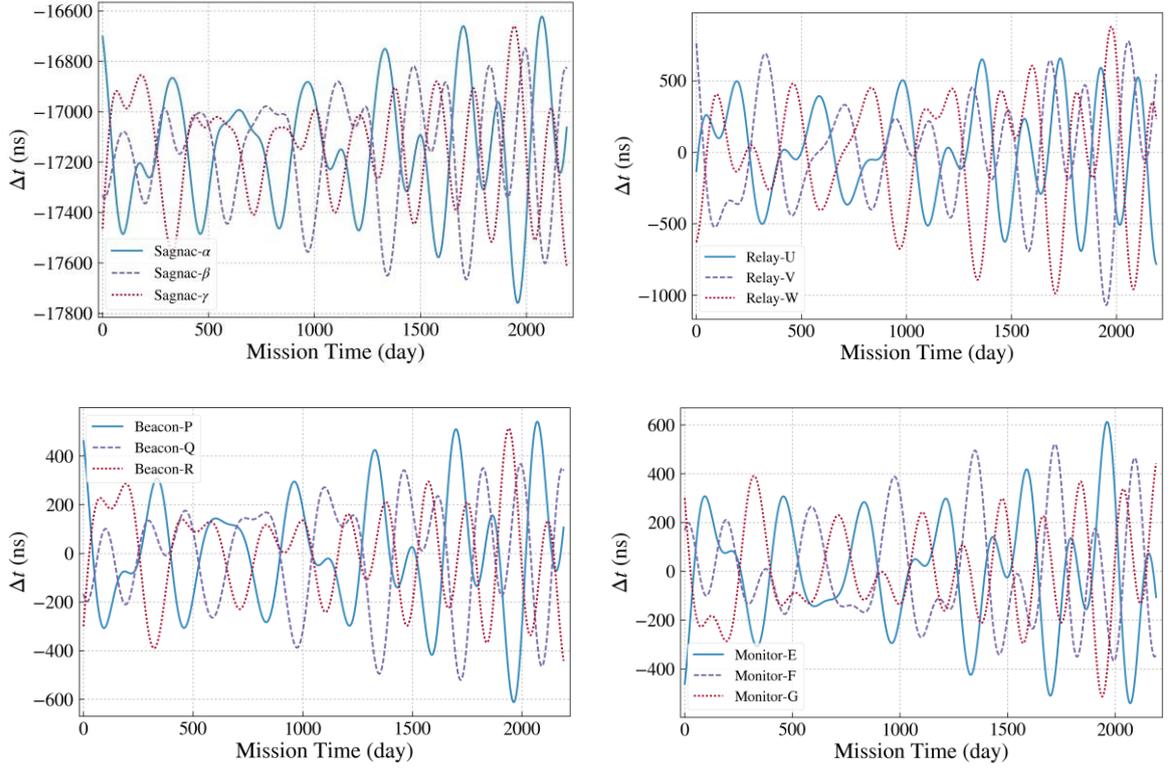

**Figure 10.** The optical path length differences vs time epoch for Sagnac *α, β & γ;* Relay U, V & W; Beacon P, Q & R; Monitor E, F & G TDIs for TAIJI.

### 3.3. Numerical results of the first-generation TDI for other LISA-like missions of various arm lengths

### 3.3.1. Unequal-arm Michelson X, Y & Z TDIs and its sum X+Y+Z for other LISA-like missions of various arm lengths

In this subsection, we draw the optical path length differences vs. time epoch for the Unequal-arm Michelson X, Y, & Z TDIs and their sum X + Y + Z for the LISA-like missions of arm length 1 Gm, 2 Gm, 4 Gm, 5 Gm and 6 Gm in figure 11.

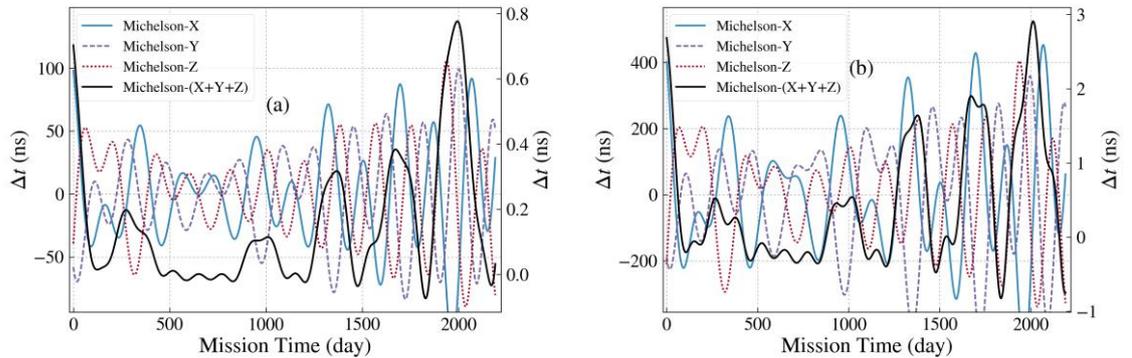

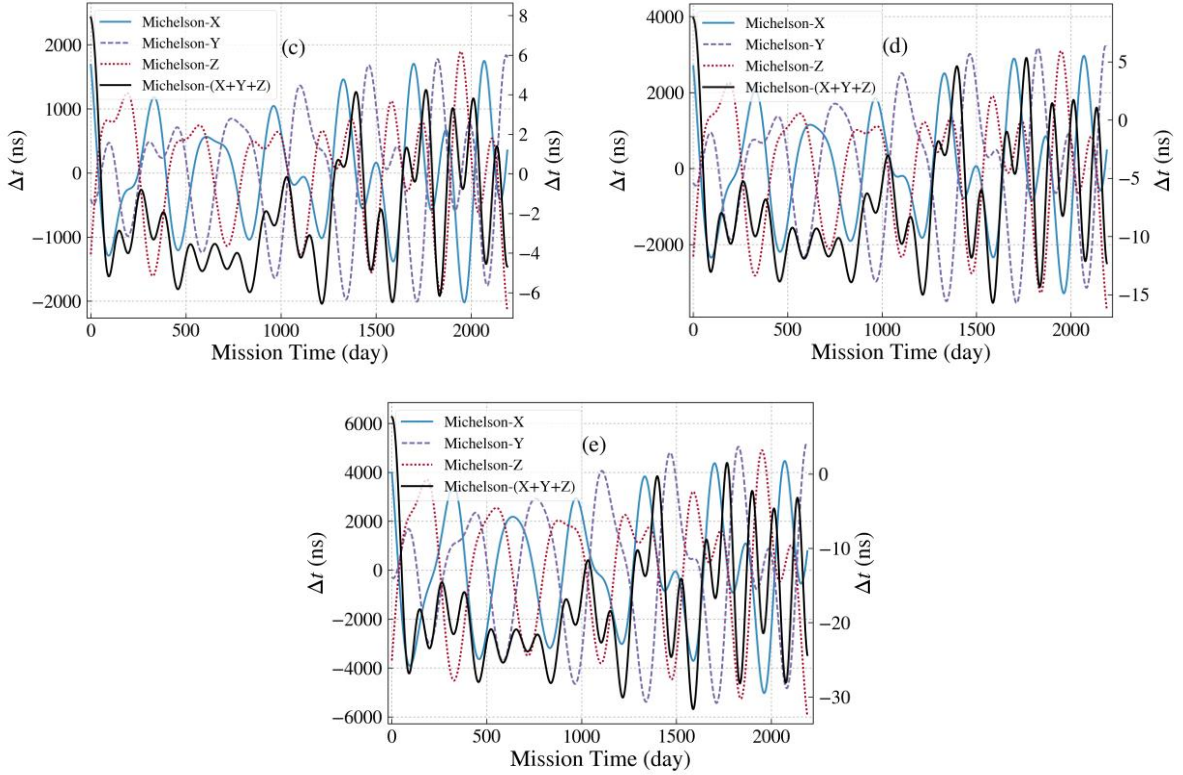

**Figure 11.** The optical path length differences for unequal-arm Michelson X, Y, & Z TDIs (units on the left of each figure) and their sum X + Y + Z (units on the right of each figure) for LISA-like missions of arm length (a) 1 Gm, (b) 2 Gm, (c) 4 Gm, (d) 5 Gm and (e) 6 Gm respectively. There is a clear cancellation of optical path length differences by 2 orders of magnitudes in the TDI sums.

*3.3.2. Sagnac α, β & γ; Relay U, V & W; Beacon P, Q & R; Monitor E, F & G first-generation TDIs for other LISA-like missions of various arm lengths*

The optical path length differences of these TDIs for eLISA mission concept of arm length 1 Gm and for a LISA-like missions of arm length 6 Gm are shown in figure 12 and figure 13 respectively. Those for LISA-like missions of arm length 2 Gm, 4 Gm and 5 Gm are shown in figures 6-8 of the supplement.

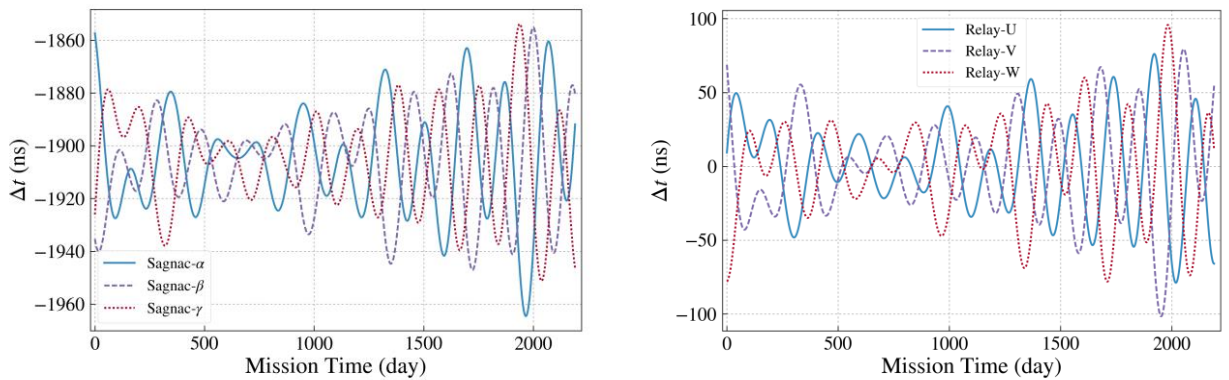

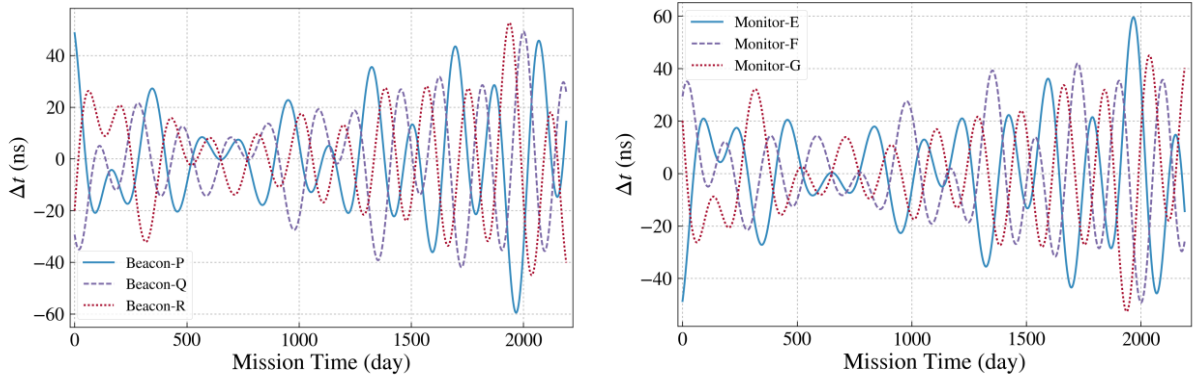

**Figure 12.** The optical path length differences vs time epoch for Sagnac *α, β & γ*; Relay U, V & W; Beacon P, Q & R; Monitor E, F & G TDIs for e-LISA/LISA-like missions of arm length 1 Gm.

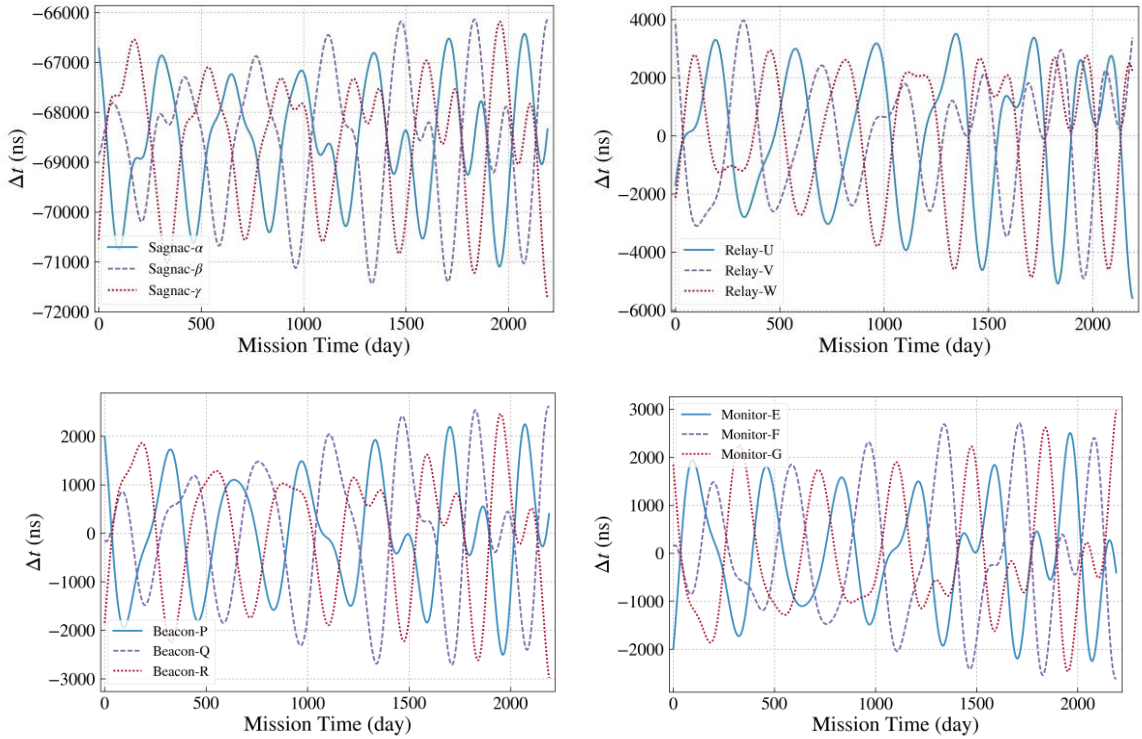

**Figure 13.** The optical path length differences vs. the time epochs for Sagnac *α, β & γ*; Relay U, V & W; Beacon P, Q & R; Monitor E, F & G *TDIs* for LISA-like missions of arm length 6 Gm.

## 4. Numerical simulation of the second-generation TDI for new LISA, TAIJI and other LISA-like missions

In this section, we do numerical simulation for the second-generation TDIs listed in subsection *1.2.2* for new LISA, TAIJI and some of the other LISA-like missions of arm length 1 Gm, 2 Gm, 4 Gm, 5 Gm and 6 Gm. We plot the optical path length differences vs. time epoch of these second-generation TDIs for LISA, TAIJI and LISA-like mission of arm length 6 Gm in subsection 4.1, 4.2 and 4.3 respectively. The plots for LISA-like missions of arm length 1 Gm, 2 Gm, 4 Gm and 5 Gm are assembled in figures 9-13 of the supplement. In table 2, we compile and compare the resulting differences for second-generation TDIs listed in subsection *1.2.2* with different arm lengths for various mission proposals -- 1 Gm (eLISA/NGO), 2 Gm (an NGO-LISA-type mission with this nominal arm length), 2.5 Gm (new LISA), 3 Gm (TAIJI), 4 Gm, 5 Gm (original LISA), 6 Gm, and ASTROD-GW.

From the last diagram in figure 14 (figure 16, or figure 18), we noticed that, the accuracy of the TDI calculation should be better than 1 μm (3.3 fs) for the path difference.

## 4.1. Numerical results of the second-generation TDIs listed in 1.2.2 for new LISA

### 4.1.1. [ab, ba], [cd, dc] and [ef, ef] (n = 1) TDI configurations, and [$a^2b^2$, $b^2a^2$], [abab, baba], [$ab^2a$, $ba^2b$], [$c^2d^2$, $d^2c^2$], [cdcd, dcdc], [$cd^2c$, $dc^2d$], [$e^2f^2$, $f^2e^2$], [efef, fefe] and [$ef^2e$, $fe^2f$] (n = 2) TDI configurations for new LISA

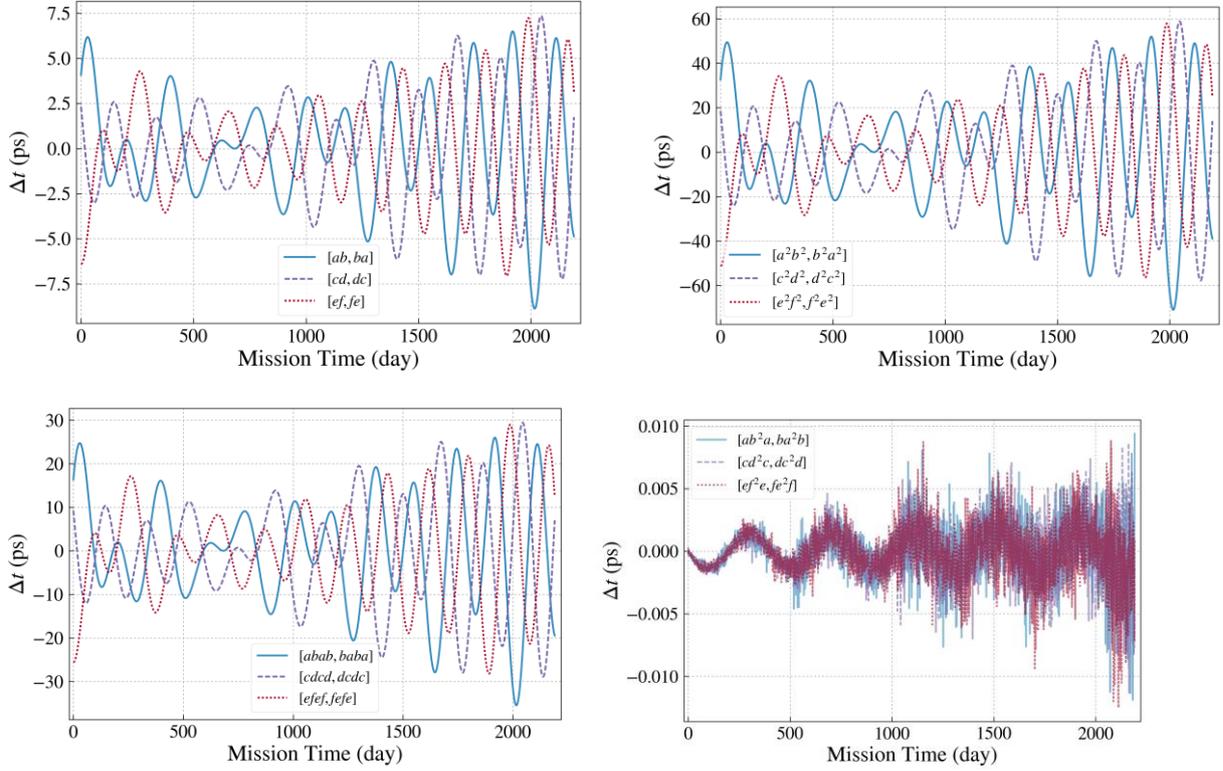

**Figure 14.** The difference of two optical path lengths vs. time epochs for [*ab*, *ba*], [*cd*, *dc*] and [*ef*, *ef*] TDI configurations (n=1), and for all n = 2 TDI configurations for new LISA.

### 4.1.2. Sagnac-type α2, β2, and γ2 TDIs for new LISA

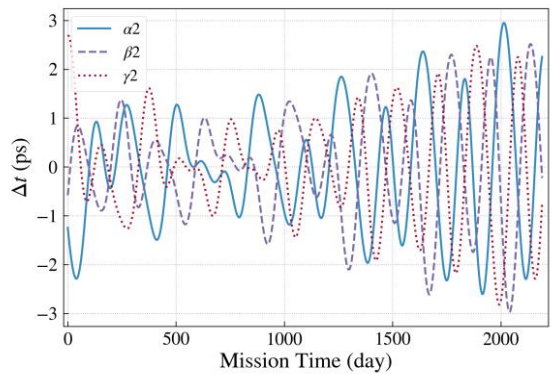

**Figure 15.** The difference of two optical path lengths vs. time epochs for Sagnac-type *α2, β2,* and *γ2* TDIs for new LISA.

## 4.2. Numerical results of the second-generation TDIs listed in subsection 1.2.2 for TAIJI

### 4.2.1. [ab, ba], [cd, dc] and [ef, ef] (n = 1) TDI configurations, and [$a^2b^2$, $b^2a^2$], [abab, baba], [$ab^2a$, $ba^2b$], [$c^2d^2$, $d^2c^2$], [cdcd, dcdc], [$cd^2c$, $dc^2d$], [$e^2f^2$, $f^2e^2$], [efef, fefe] and [$ef^2e$, $fe^2f$] TDI configurations (n = 2) for TAIJI

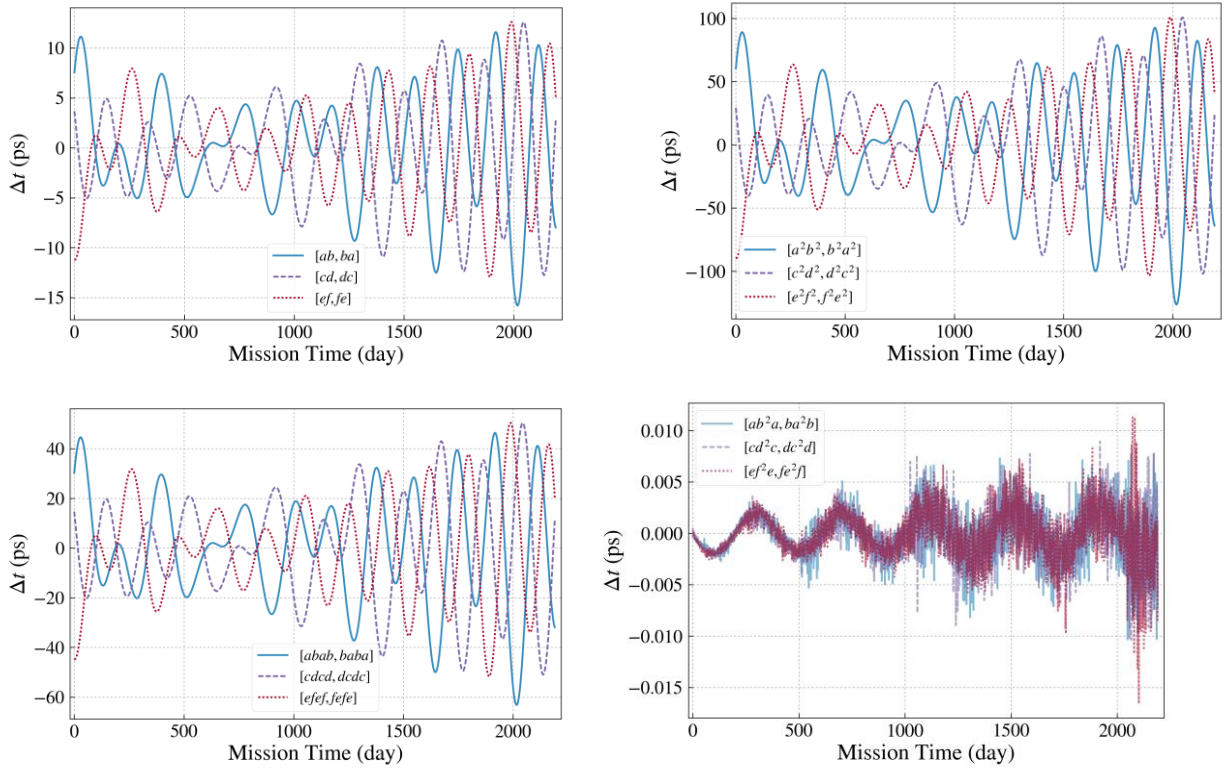

**Figure 16.** The difference of two optical path lengths vs. time epochs for [*ab, ba*], [*cd, dc*] and [*ef, ef*] TDI configurations (n=1), and for all n = 2 TDI configurations for TAIJI.

*4.2.2. Sagnac-type α2, β2, and γ2 TDIs for TAIJI*

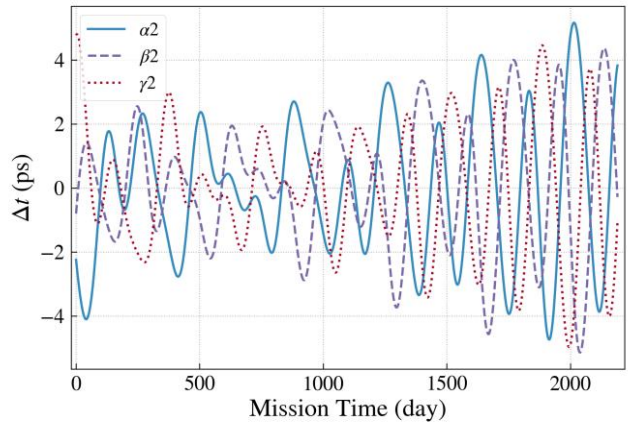

**Figure 17.** The difference of two optical path lengths vs. time epochs for Sagnac-type *α2, β2,* and *γ2* TDIs for TAIJI.

*4.3. Numerical results of the second-generation TDIs listed in subsection 1.2.2 for LISA-like mission of arm length 6 Gm*

*4.3.1. [ab, ba], [cd, dc] and [ef, ef] (n = 1) TDI configurations, and [$a^2b^2$, $b^2a^2$], [abab, baba], [$ab^2a$, $ba^2b$], [$c^2d^2$, $d^2c^2$], [cdcd, dcdc], [$cd^2c$, $dc^2d$], [$e^2f^2$, $f^2e^2$], [efef, fefe]; [$ef^2e$, $fe^2f$] (n = 2) TDI configurations for LISA-like mission of arm length 6 Gm*

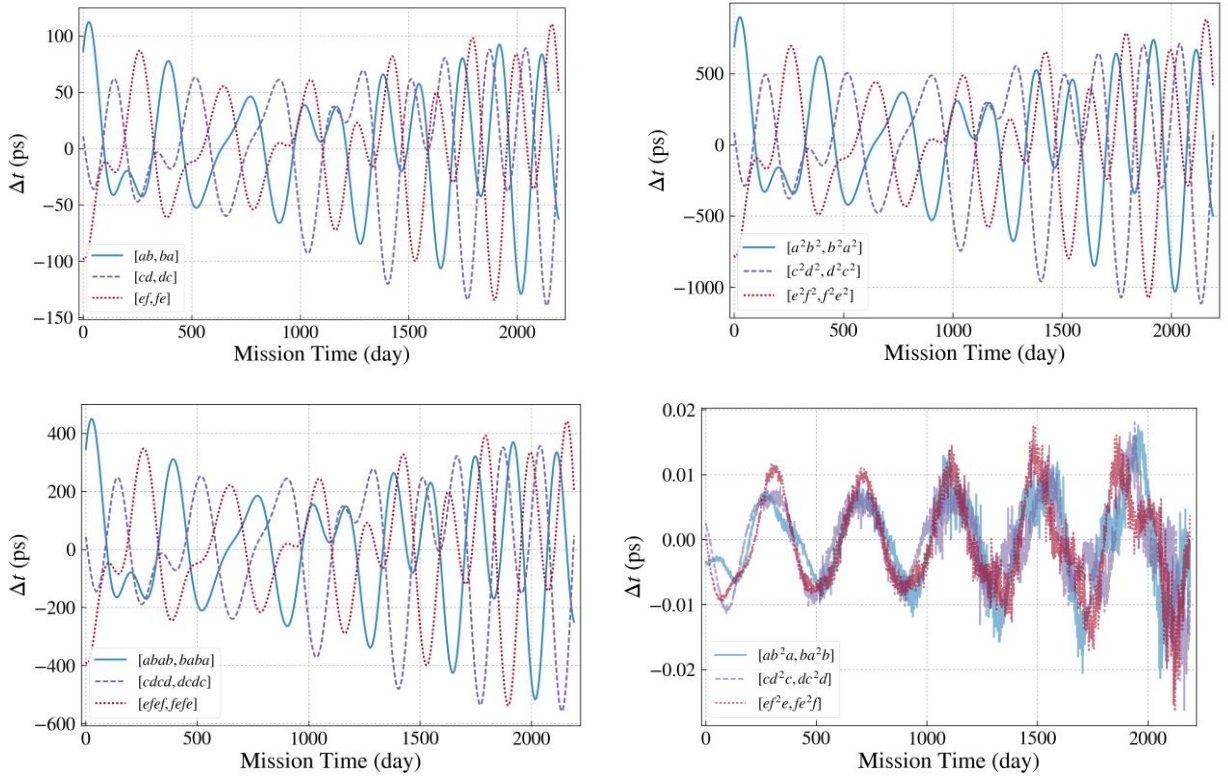

**Figure 18.** The difference of two optical path lengths vs. time epochs for [*ab, ba*], [*cd, dc*] and [*ef, ef*] TDI configurations (n=1), and for all n = 2 TDI configurations for LISA-like mission of arm length 6 Gm.

*4.3.2. Sagnac-Type α2, β2, and γ2 TDIs for LISA-like mission of arm length 6 Gm*

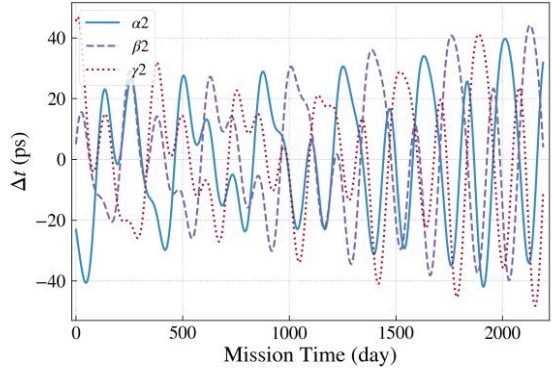

**Figure 19.** The difference of two optical path lengths vs. time epochs for Sagnac-type *α2, β2,* and *γ2* TDIs for LISA-like mission of arm length 6 Gm.

**5. Discussion and outlook**

We have optimized a set of 6-year (2200-day) new LISA, TAIJI and other LISA-like mission orbits numerically using ephemeris framework starting at March 22, 2028. The line-of-sight Doppler velocities basically satisfy the respective requirements. The maximum magnitude and rms magnitude together with the upper-limit requirement vs. arm length are plotted in figure 20. It is clear that the upper-limit follows linear trend closely. We calculate optical path length differences of various first-generation and second-generation TDIs for these mission orbits and compile them in table 1 and table 2 together with those of ASTROD-GW for easy comparison. We list the presently recommended requirement in the last row of the tables. From the tables and the figures in sections 3 and 4, we have the following:

i) All the first-generation TDIs violate their respective requirements. For eLISA/NGO of arm length 1 Gm, the deviation for X, Y, and Z TDIs could be up to a factor of 3.6; for the case of 2 Gm arm length, a factor of 6.8; for the new LISA case of 2.5 Gm arm length, a factor of 9.6; for the TAIJI case of 3 Gm arm length, a factor of 12.5; for the case of 4 Gm arm length, a factor of 15.2; for the original LISA of 5 Gm, a factor of 22.3; for the case of 6 Gm arm length, a factor 29.9. *If X, Y, and Z TDIs are used for the GW analysis, either the TDI requirement needs to be relaxed by the same factor or laser frequency stability requirement needs to be strengthened by the same factor.* In upper part of figure 21, we plot the largest value of the rms averages of X, Y and Z TDIs vs. arm length. It looks like a power law dependence on the arm length. In the lower part of figure 21, we log-log plot of the largest of the rms averages of X, Y and Z TDIs vs arm length with ASTROD-GW on the same plot. Indeed, for LISA-like missions, it is nearly linear on this plot with slope (power index) about 2.3. The point on this diagram for ASTROD-GW is much lower than the linear extrapolation of this trend. This could because that the ASTROD-GW S/Cs are near L4, L5 stable points and L3 quasi-stable points (with a time scale of instability of about 50 years).

(ii) All the second-generation TDIs in table 2 for eLISA/NGO of arm length 1 Gm, for the case of 2 Gm arm length and for the new LISA case of 2.5 Gm arm length, for the TAIJI of 3.0 Gm arm length, for the case of 4 Gm arm length, for the original LISA of 5 Gm, and for the case of 6 Gm arm length satisfy their respective requirements with good margins. Nevertheless, the GW response and sensitivity needs to be calculated for the GW data analysis.

(iii) Experimental demonstration of TDI in laboratory for LISA has been implemented in 2010-2012 (Vine *et al*, 2010; Mirtyk *et al*, 2012). eLISA and the original ASTROD-GW TDI requirement are based on LISA requirement, and hence also demonstrated. With the present pace of development in laser technology, the laser frequency noise requirement is expected to be able to compensate for 1-2 order TDI requirement relaxation in 10 years. For new LISA and TAIJI, the X, Y and Z TDIs could also be considered.

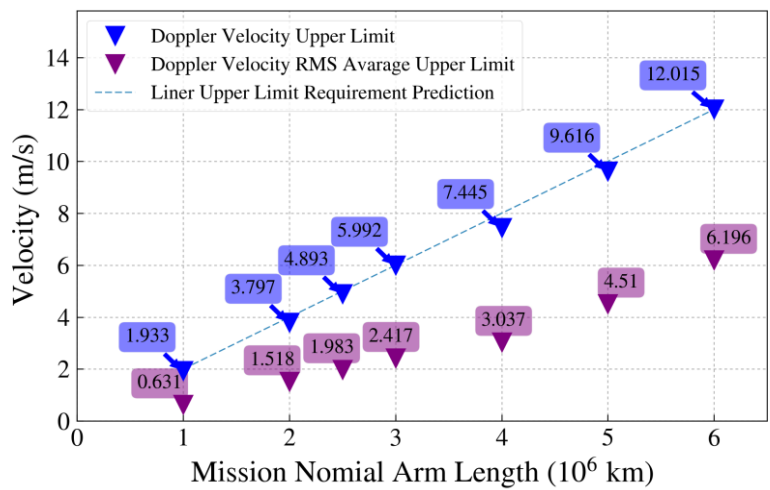

**Figure 20.** Line of sight Doppler velocities (maximum and rms average in 2200 days) vs. arm length.

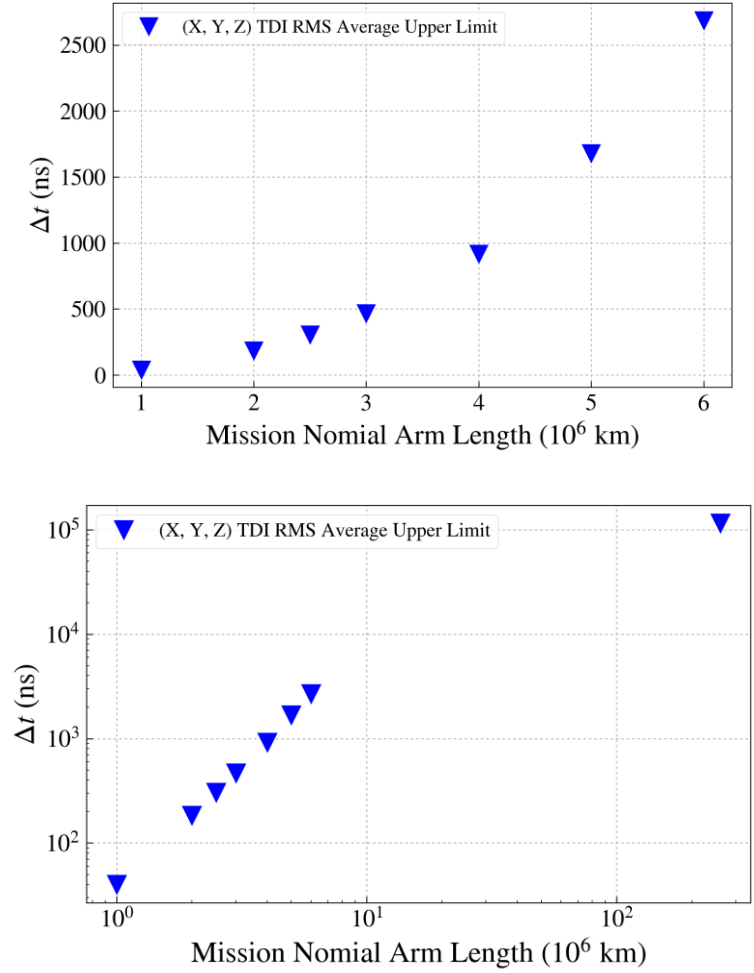

**Figure 21.** (upper diagram) The largest of the rms averages of X, Y and Z TDIs vs. arm length. (lower diagram) Log-log plot of the largest of the rms averages of X, Y and Z TDIs vs arm length with ASTROD-GW on the same plot.

**Appendix. CGC ephemeris framework**

In 1998, we started orbit simulation and parameter determination for ASTROD (Chiou and Ni, 2000a, 2000b), and worked out a post-Newtonian ephemeris of the Sun, the major planets and 3 biggest asteroids including the solar quadrupole moment. This working ephemeris was termed as CGC 1 (CGC: Center for Gravitation and Cosmology). For an improved ephemeris framework, we considered all known 492 asteroids with diameter greater than 65 km to obtain the CGC 2 ephemeris, and calculated the perturbations due to these 492 asteroids on the ASTROD spacecraft (Tang and Ni, 2000, 2002). In building the CGC ephemeris framework, we use the post-Newtonian barycentric metric and equations of motion as derived in Brumberg (1991) for solar system bodies with PPN (Parametrized Post-Newtonian) parameters $\beta$, $\gamma$. In solving a problem, one may use any coordinate system. However, in our ephemeris, we just use the equations in Brumberg (1991) with gauge parameters $\alpha = \nu = 0$ that corresponds to the harmonic gauge adopted by the 2000 IAU resolution (Soffel, 2003).

In our first optimization of ASTROD-GW orbits (Men *et al*, 2009, 2010), we used the CGC 2.5 ephemeris in which only 3 biggest minor planets are taken into account, but the Earth's precession and nutation are added; the solar quadratic zonal harmonic and the Earth's quadratic to quartic zonal harmonic are also included.

In our recent orbit simulation of the ASTROD I proposed to ESA (Braxmaier *et al*, 2012) and in our studies of TDIs for LISA (Dhurandhar, Ni and Wang, 2013) and for ASTROD-GW (Wang and Ni, 2011, 2012, 2013; Wang, 2011), we added the perturbation of additional 349 asteroids to the CGC 2.5 ephemeris and called it the CGC 2.7 ephemeris. (The difference between the CGC 2.7 ephemeris and the CGC 2 ephemeris is that we have 352 asteroids instead of 492 asteroids) For more discussions on the CGC 2.7 ephemeris, please see Wang and Ni (2011, 2012).

In the CGC 2.7.1 ephemeris framework, we pick up 340 asteroids besides the Ceres, Pallas, and Vesta from the Lowell database (The Asteroid Orbital Elements Database, ftp://ftp.lowell.edu/pub/elgb/astorb.html). The masses of 340 asteroids are given by DE430/DE431 (Folkner et al 2014) instead of estimating the masses based on the classification in CGC 2.7. The orbit elements of these asteroids are also updated from the Lowell database.

**Acknowledgements**
GW receives funding in support of his work leading to these results from the People Programme (Marie Curie Actions) of the European Union's Seventh Framework Programme FP7/2007-2013/ (PEOPLE-2013-ITN) under REA grant agreement n° [606176]. It reflects only the authors' view and the Union is not liable for any use that may be made of the information contained therein. WTN would like to thank Rong-Gen Cai for supporting his visit at KITPC while finishing this work.

# Supplement for TDI simulation

Gang Wang and Wei-Tou Ni

E-mail: gwanggw@gmail.com, weitou@gmail.com

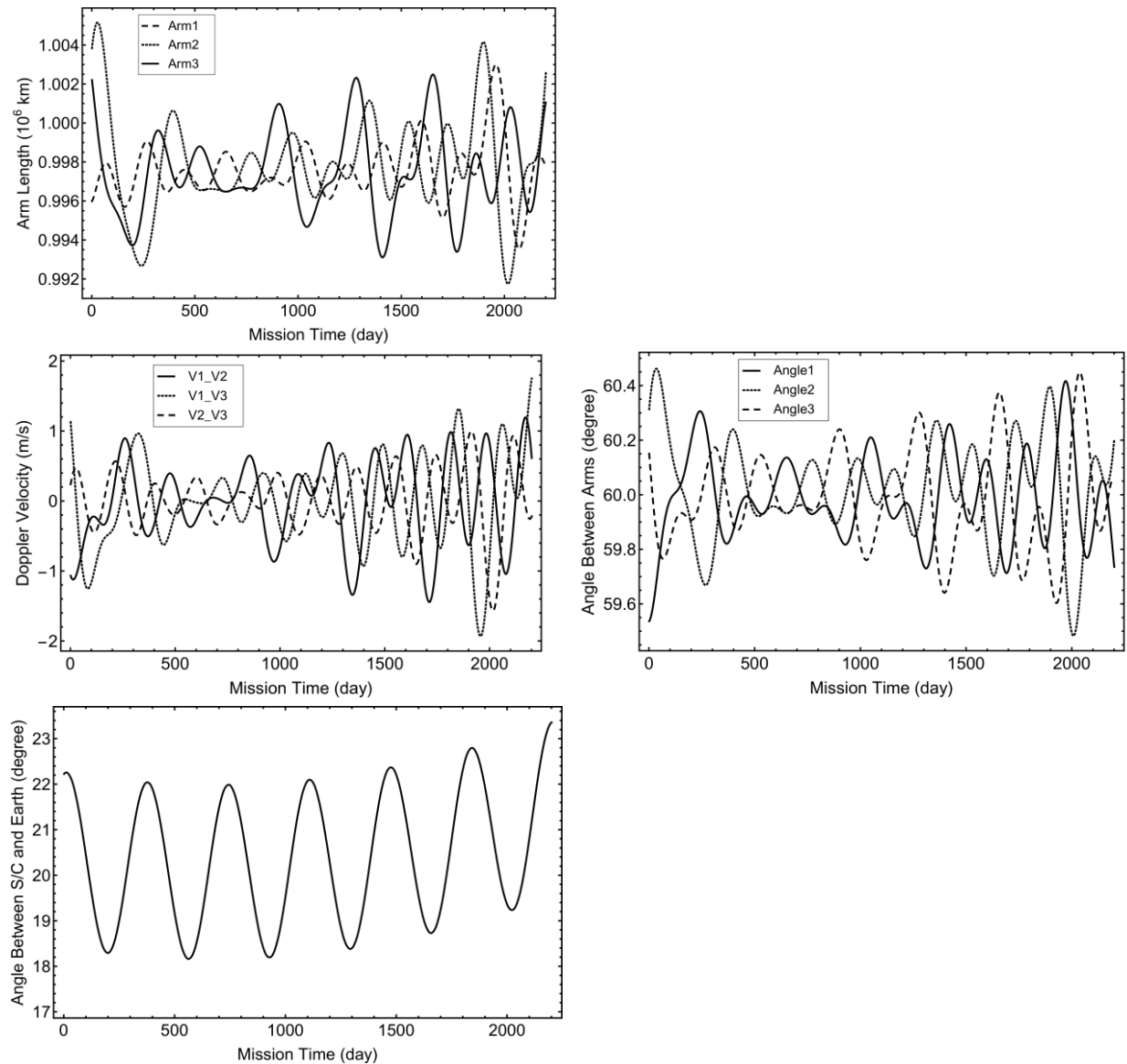

**Figure 1.** Variations of the arm lengths, the Doppler velocities, the formation angles and the angle between barycentre of S/C and Earth in 2200 days for the 1 Gm LISA- like S/C configuration with initial conditions given in Table 5.

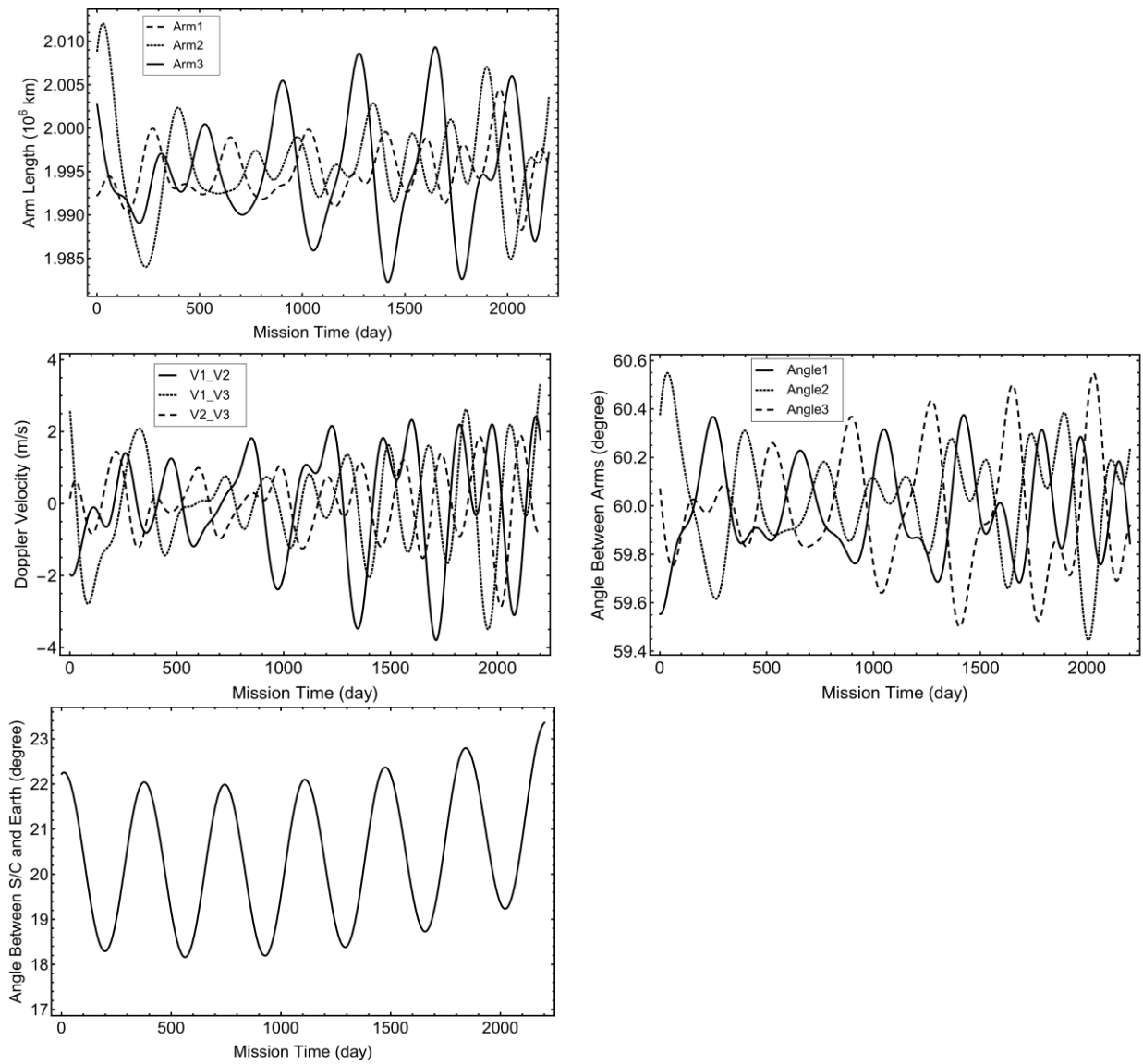

**Figure 2**. Variations of the arm lengths, the Doppler velocities, the formation angles and the angle between barycentre of S/C and Earth in 2200 days for the 2 Gm LISA-like S/C configuration with initial conditions given in Table 5.

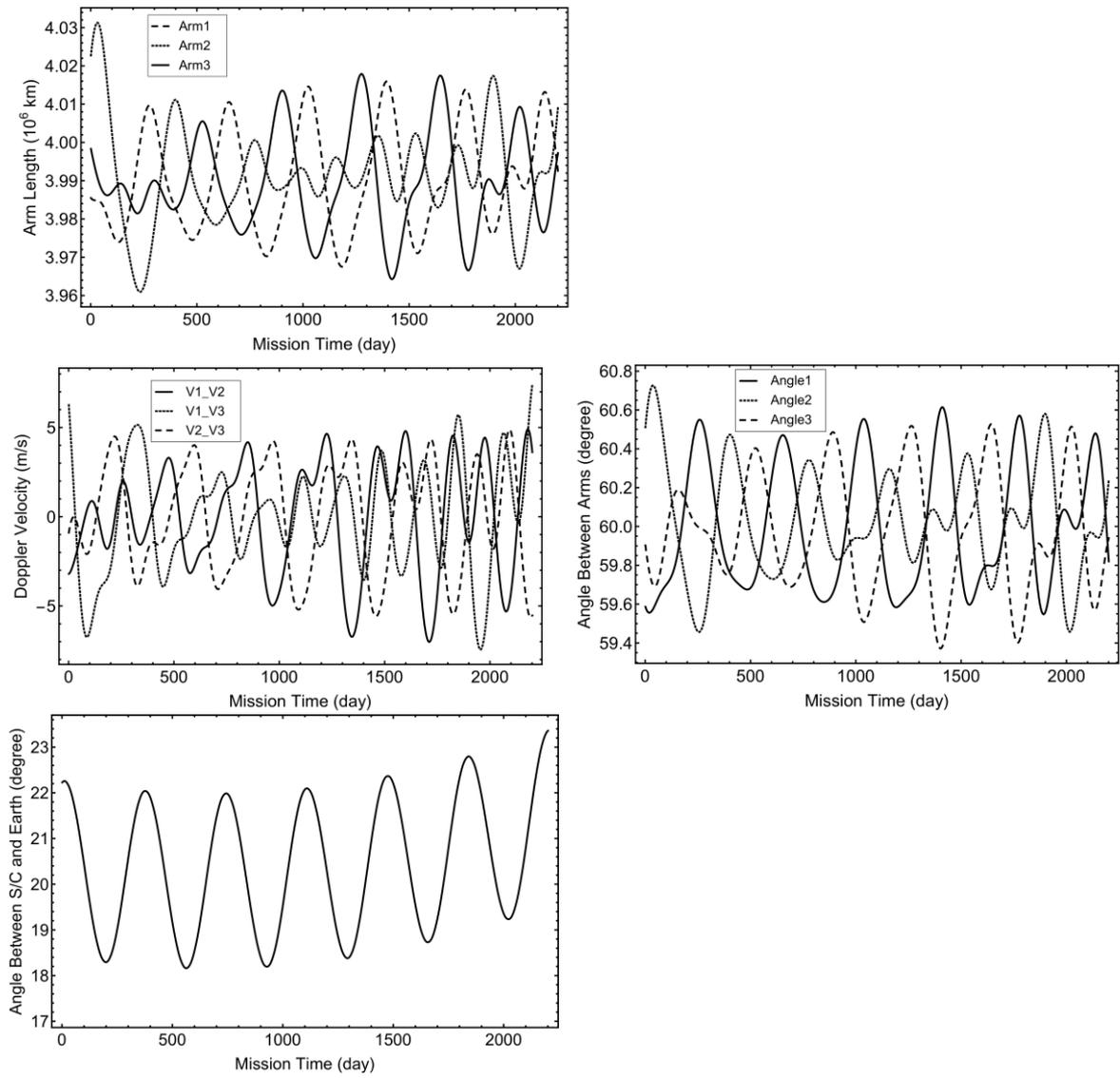

**Figure 3**. Variations of the arm lengths, the Doppler velocities, the formation angles and the angle between barycentre of S/C and Earth in 2200 days for the 4 Gm LISA-like S/C configuration with initial conditions given in Table 5.

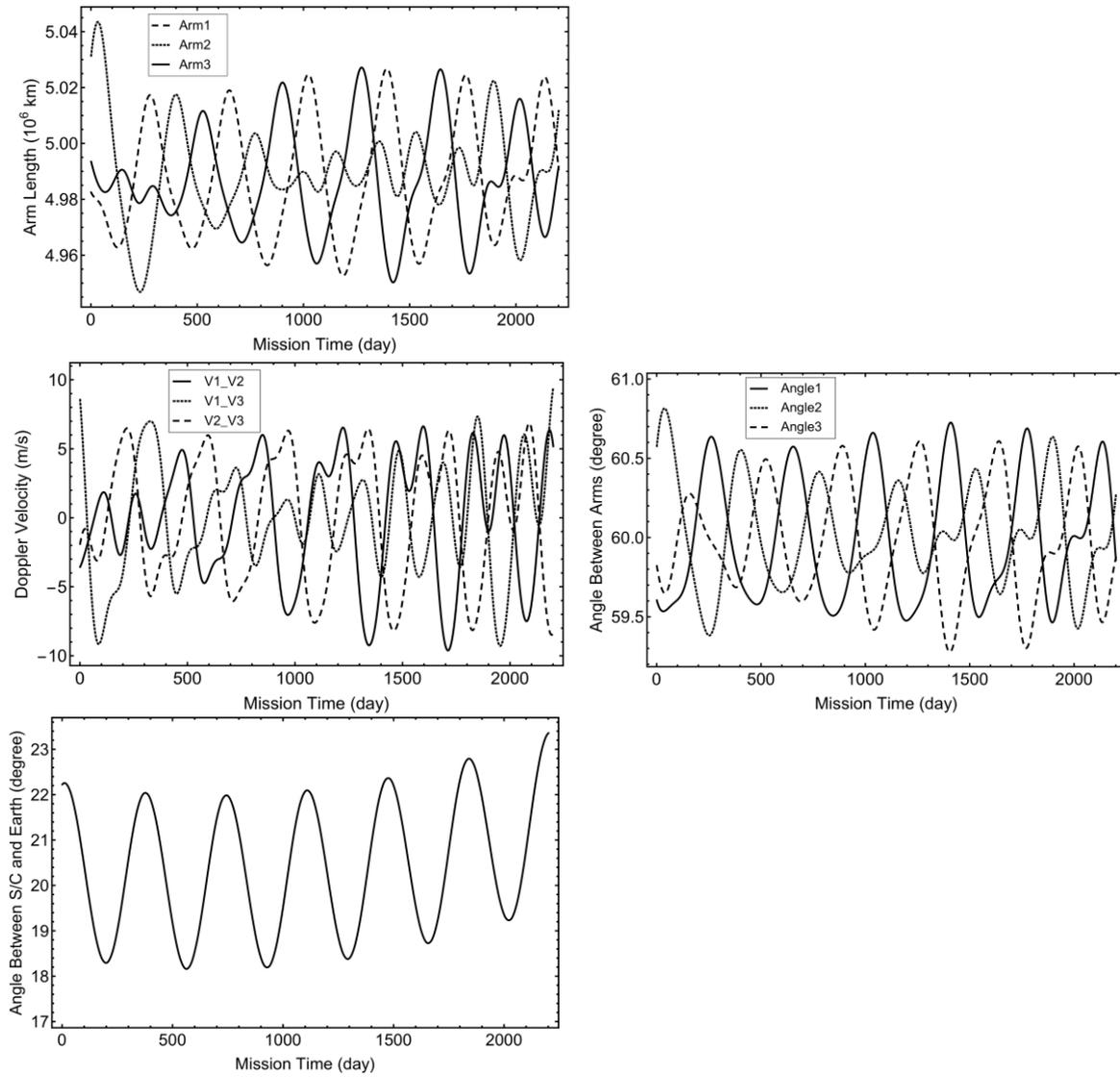

**Figure 4**. Variations of the arm lengths, the Doppler velocities, the formation angles and the angle between barycentre of S/C and Earth in 2200 days for the 5 Gm LISA-like S/C configuration with initial conditions given in Table 5.

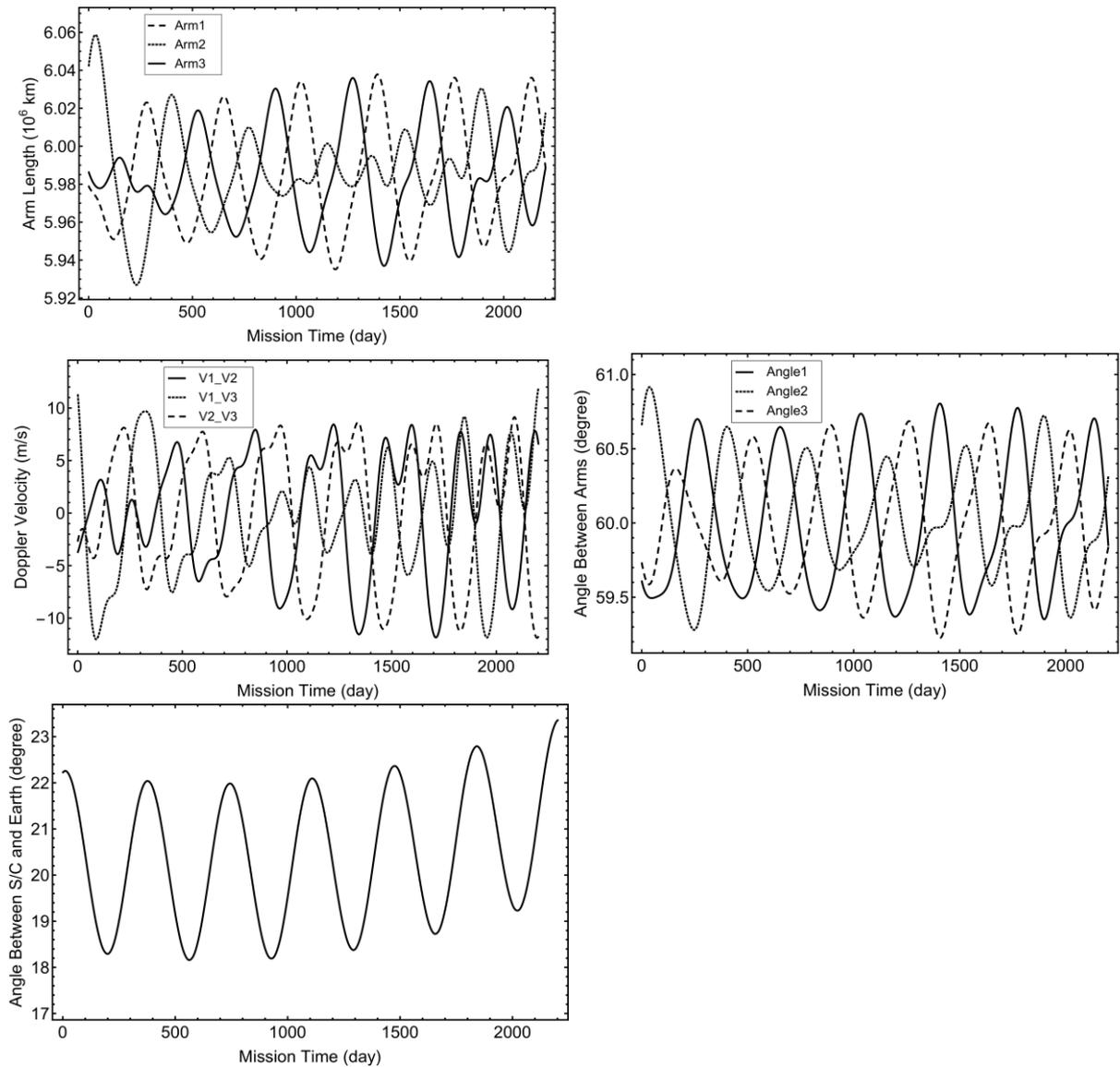

**Figure 5**. Variations of the arm lengths, the Doppler velocities, the formation angles and the angle between barycentre of S/C and Earth in 2200 days for the 6 Gm LISA- like S/C configuration with initial conditions given in Table 5.

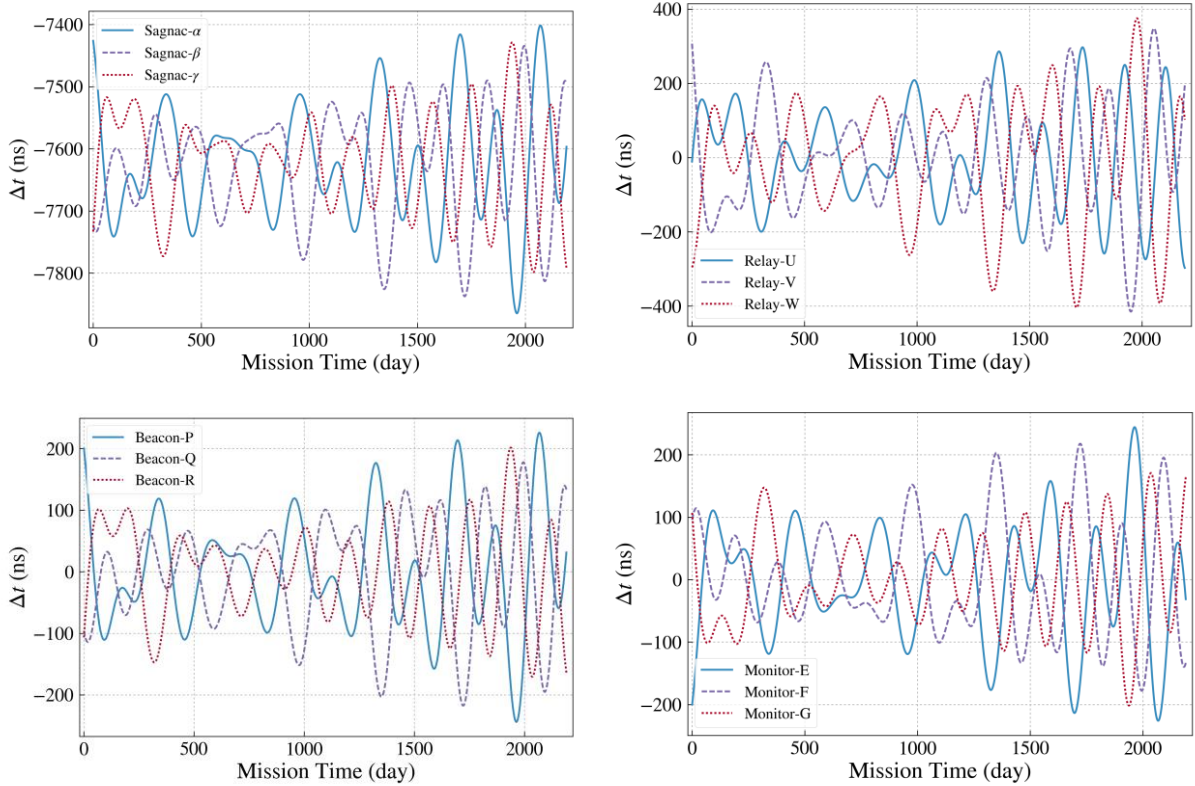

**Figure 6**. The optical path length differences vs time epoch for Sagnac (α, β, γ); Relay (U, V, W); Beacon (P, Q, R); Monitor (E, F, G) TDIs for LISA-like missions of arm length 2 Gm.

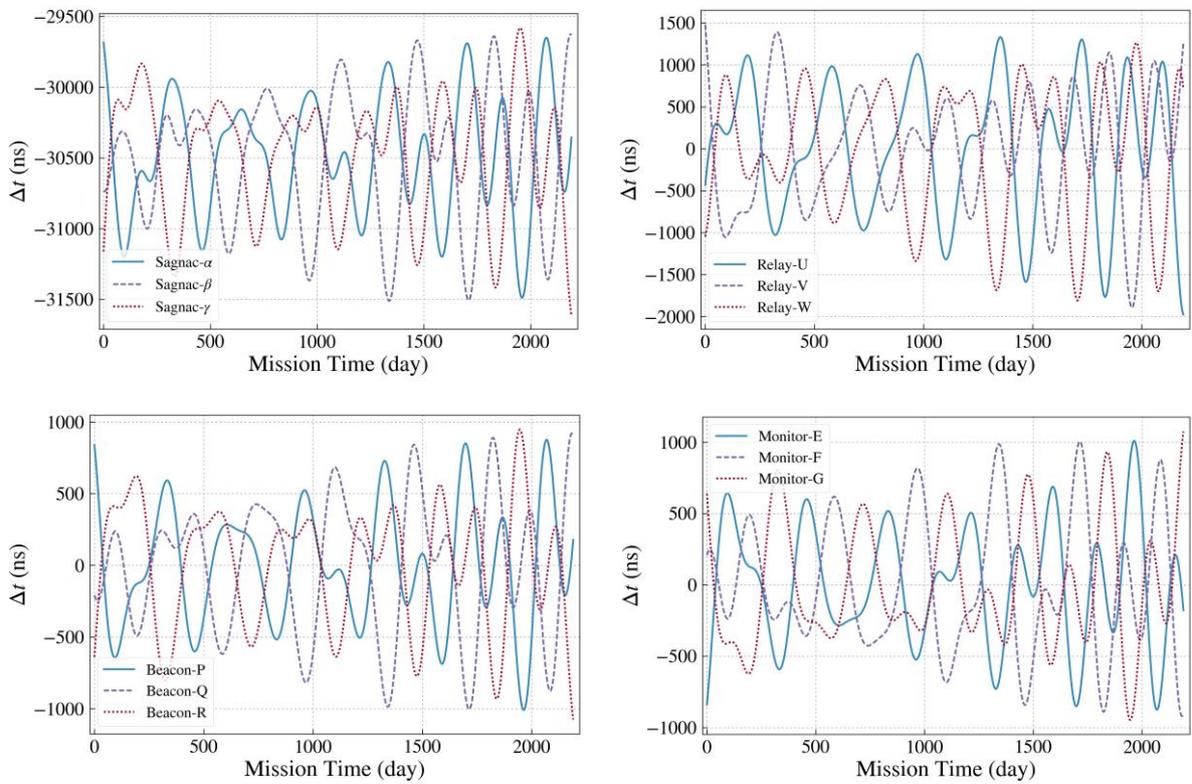

**Figure 7**. The optical path length differences vs time epoch for Sagnac (α, β, γ); Relay (U, V, W); Beacon (P, Q, R); Monitor (E, F, G) TDIs for LISA-like missions of arm length 4 Gm.

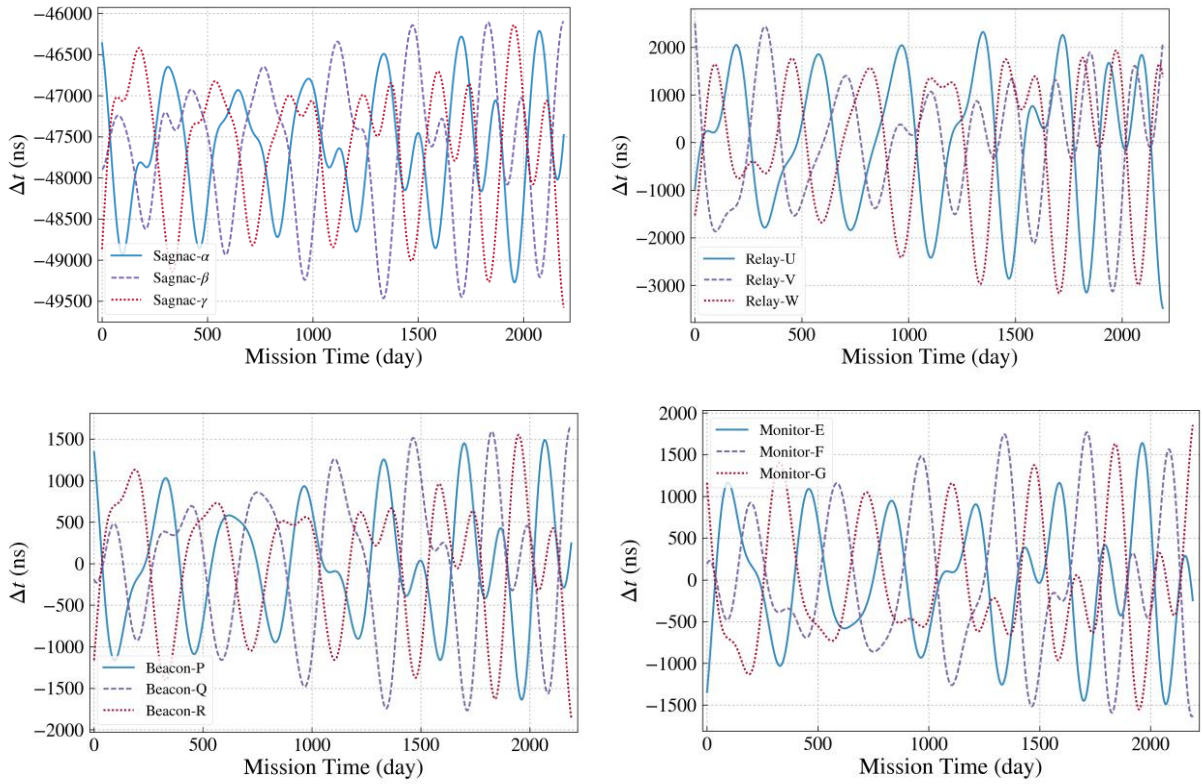

**Figure 8**. The optical path length differences vs time epoch for Sagnac (α, β, γ); Relay (U, V, W); Beacon (P, Q, R); Monitor (E, F, G) TDIs for LISA-like missions of arm length 5 Gm.

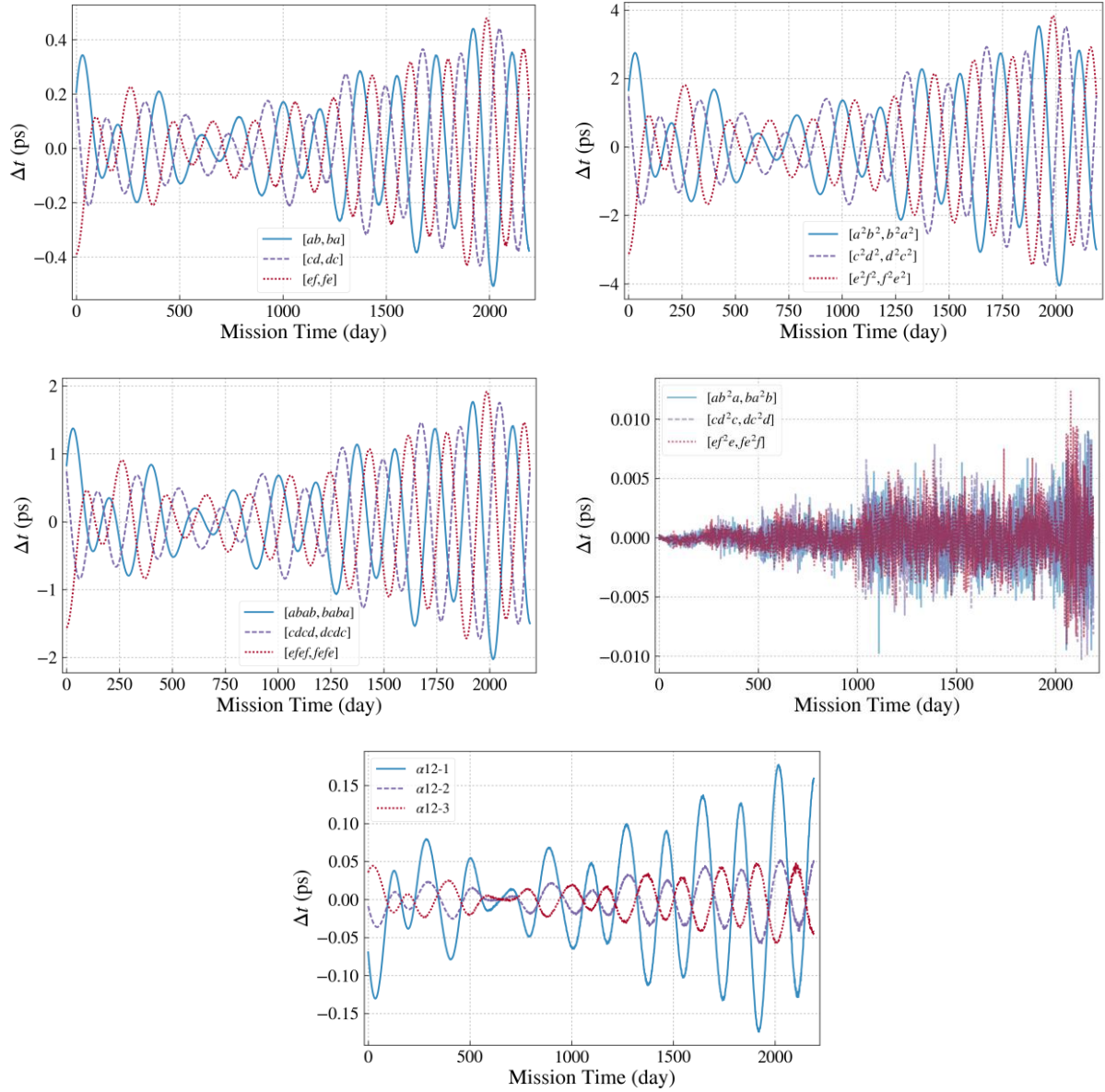

**Figure 9**. The difference of two optical path lengths versus time epochs for [ab,ba]-type TDI configurations (n=1), and for [$a^2b^2,b^2a^2$], [abab,baba] and [$ab^2a,ba^2b$] type TDI configurations (n = 2), and the α12-type for 1 Gm.

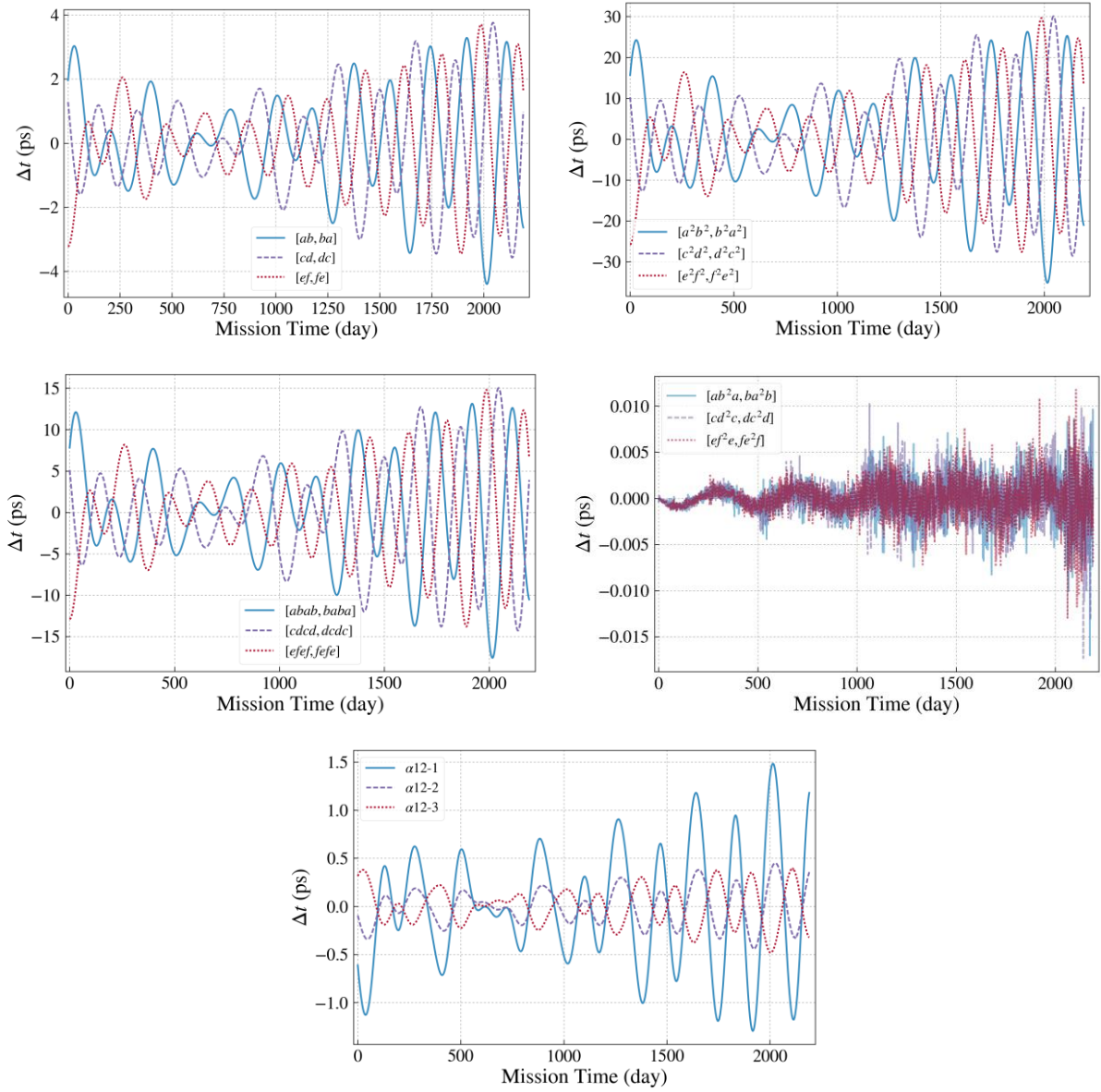

**Figure 10**. The difference of two optical path lengths versus time epochs for [ab, ba]-type TDI configurations (n=1), and for [$a^2b^2,b^2a^2$], [abab,baba] and [$ab^2a,ba^2b$] type TDI configurations (n = 2), and the α12-type for 2 Gm.

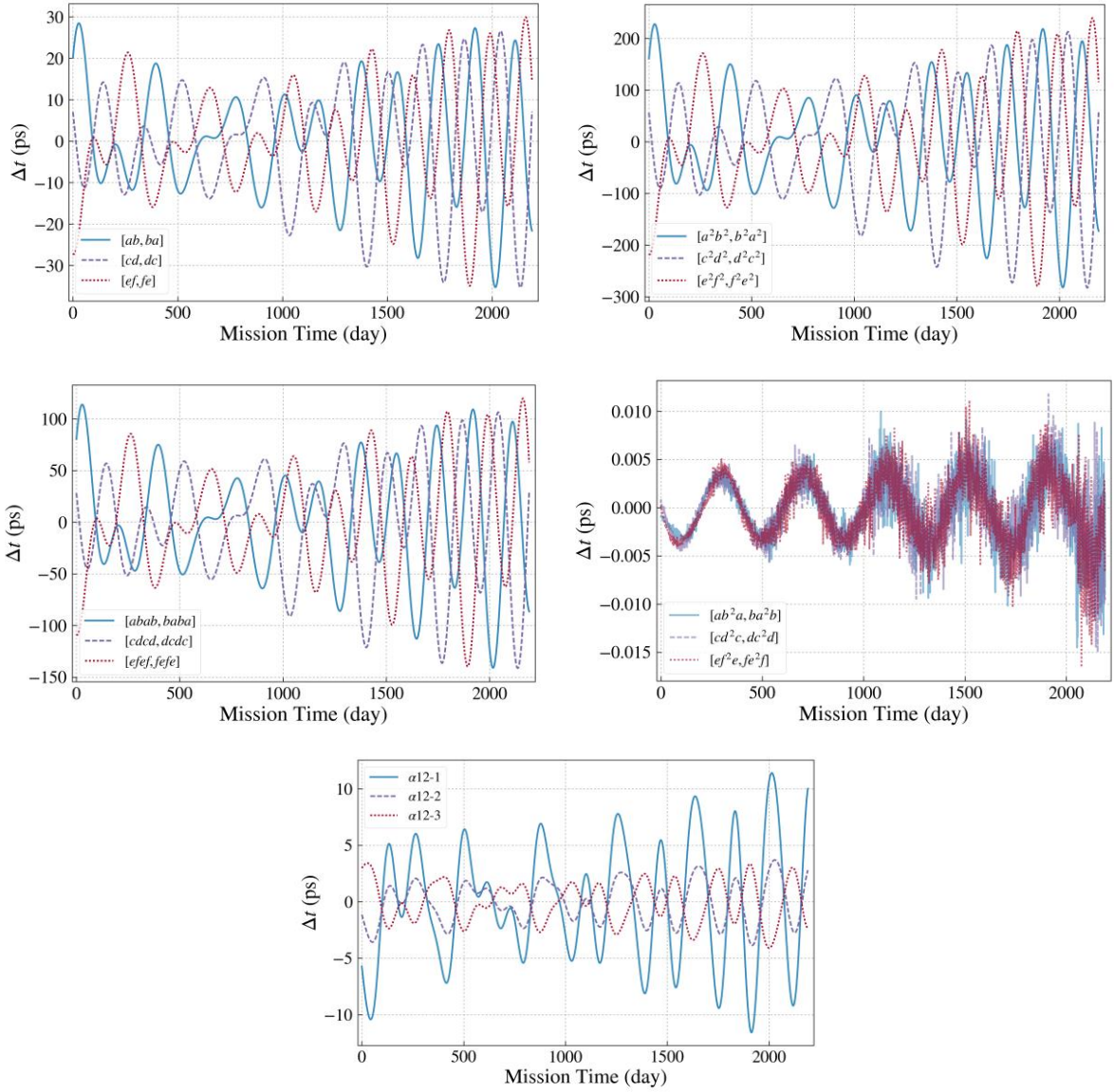

**Figure 11**. The difference of two optical path lengths versus time epochs for [ab, ba]-type TDI configurations (n=1), and for [$a^2b^2,b^2a^2$], [abab,baba] and [$ab^2a,ba^2b$] type TDI configurations (n = 2), and the α12-type for 4 Gm.

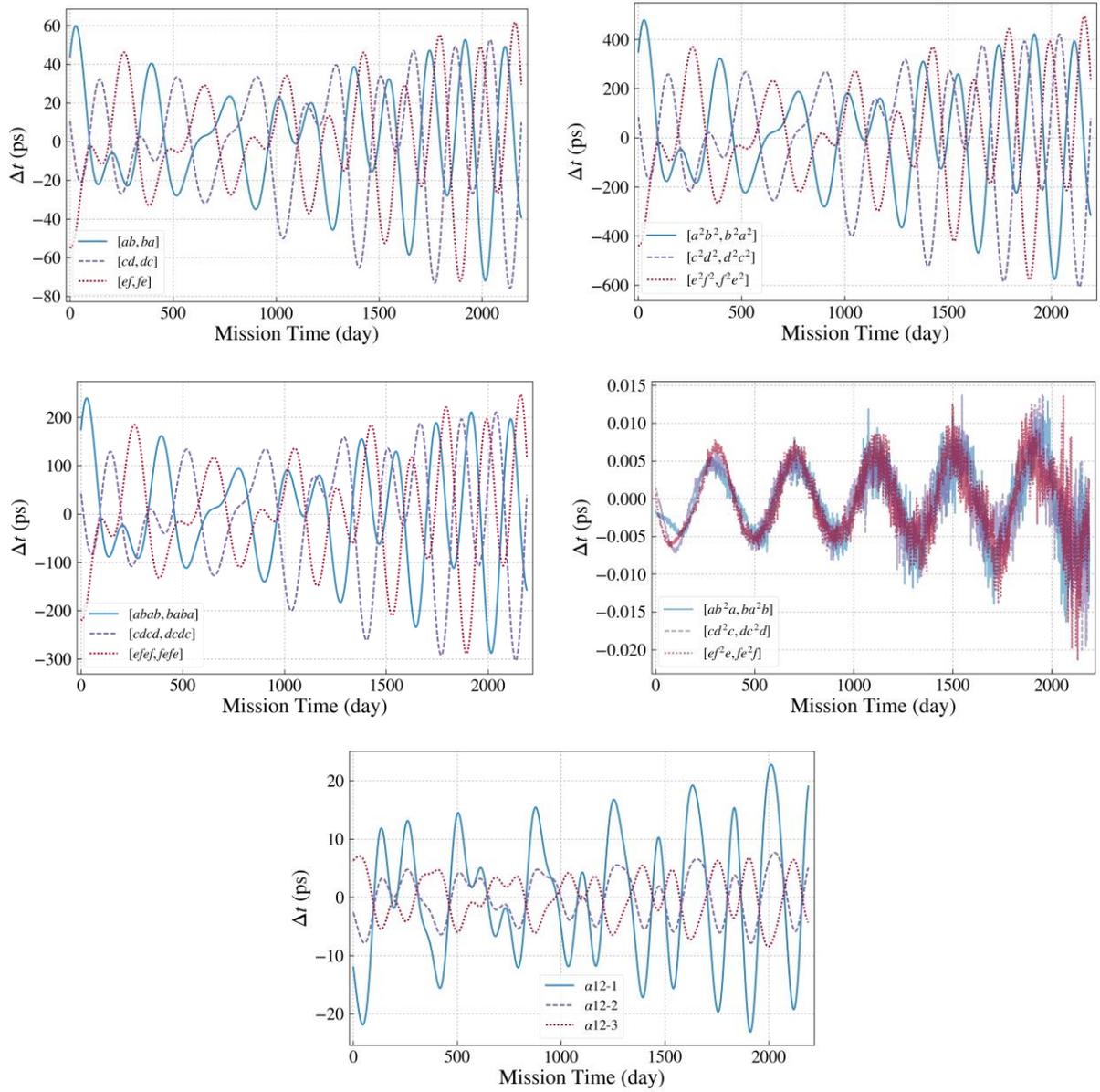

**Figure 12**. The difference of two optical path lengths versus time epochs for [ab, ba]-type TDI configurations (n=1), and for [$a^2b^2$,$b^2a^2$], [abab,baba] and [$ab^2a$,$ba^2b$] type TDI configurations (n = 2), and the α12-type for 5 Gm.

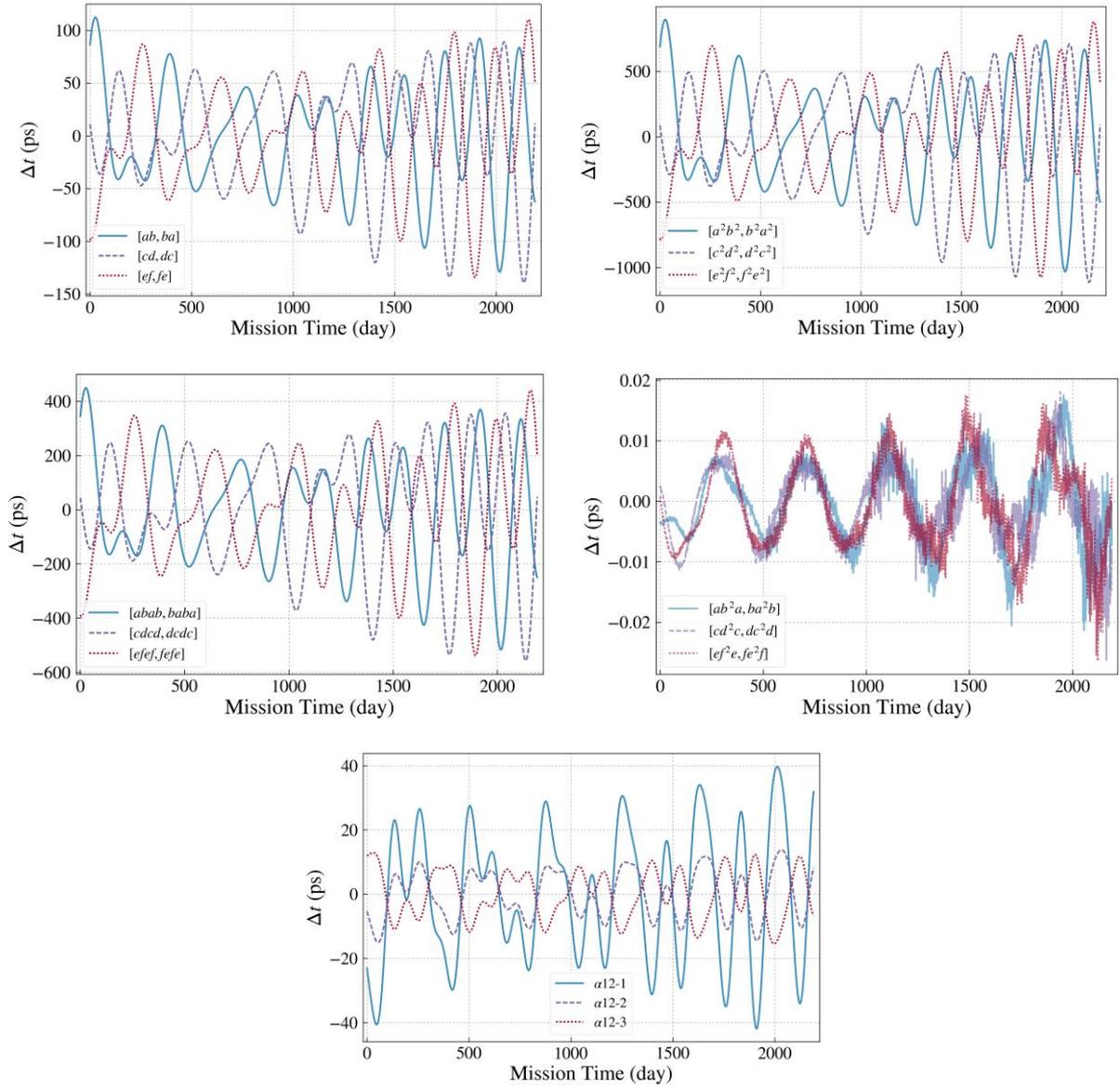

**Figure 13**. The difference of two optical path lengths versus time epochs for [ab, ba]-type TDI configurations (n=1), and for [$a^2b^2,b^2a^2$], [abab,baba] and [$ab^2a,ba^2b$] type TDI configurations (n = 2), and the α12-type for 6 Gm.